\documentclass[iop]{emulateapj} 
\slugcomment{Draft of \today}

\usepackage{amsmath} \usepackage{epstopdf} \usepackage{graphicx}
\usepackage[table]{xcolor} \usepackage{mathrsfs} \usepackage{hyperref}
\usepackage{bm} \usepackage{verbatim} \usepackage{bigints} \usepackage{pbox} \usepackage{enumitem}

\newcommand{\Mrem}{M_{\rm rem}}
\newcommand{\Mto}{M_{\rm to}}
\newcommand{\frem}{f_{\rm rem}}
\newcommand{\Frec}{F_{\rm rec}}
\newcommand{\Fdotrec}{\dot{F}_{\rm rec}}
\newcommand{\Mmin}{M_{\rm min}}
\newcommand{\Mmax}{M_{\rm max}}
\newcommand{\mdotstar}{\dot{M}_*}
\newcommand{\Zo}{Z_{\rm O}}
\newcommand{\Mo}{M_{\rm O}}

\newcommand{\MFe}{M_{\rm Fe}}
\newcommand{\mocc}{m_{\rm O}^{\rm CC}}
\newcommand{\mfecc}{m_{\rm Fe}^{\rm CC}}
\newcommand{\mfeIa}{m_{\rm Fe}^{\rm Ia}}
\newcommand{\tsf}{t_{\rm sf}}
\newcommand{\Gyr}{\,{\rm Gyr}}
\newcommand{\kpc}{\,{\rm kpc}}
\newcommand{\ofe}{[{\rm O}/{\rm Fe}]}
\newcommand{\afe}{[\alpha/{\rm Fe}]}
\newcommand{\feh}{[{\rm Fe}/{\rm H}]}
\newcommand{\Nstari}{N_{*,i}}

\newcommand{\fstari}{f_{*,i}}
\newcommand{\fstaridat}{f^{\rm data}_{*,i}}
\newcommand{\fstarimod}{f^{\rm model}_{*,i}}

\begin{document} 

\title{The Chemical Abundance Structure of the Inner Milky Way: A Signature of
``Upside-Down" Disk Formation?} 
\author{
Jenna K.~C.~Freudenburg\altaffilmark{1},
David H.~Weinberg\altaffilmark{1},
Michael R.~Hayden\altaffilmark{2},
Jon A.~Holtzman\altaffilmark{2}
}

\altaffiltext{1}{Department of Astronomy, The Ohio State University, 
140 W.  18th Ave., Columbus, OH 43210, USA 
(freudenburg.2, weinberg.21@osu.edu)}
\altaffiltext{2}{New Mexico State University, Las Cruces, NM 88003, USA
(mrhayden, holtz@nmsu.edu)}

\begin{abstract} 
We present a model for the $\afe$-$\feh$ distribution of stars in the 
inner Galaxy, $3\kpc<R<5\kpc$, measured as a function of vertical distance 
$|z|$ from the midplane by Hayden et al.\ (2015, H15). Motivated by an 
``upside-down'' scenario for thick disk formation, in which the thickness of 
the star-forming gas layer contracts as the stellar mass of the disk grows, we 
combine one-zone chemical evolution with a simple prescription in which the 
scale-height of the stellar distribution drops linearly from $z_h=0.8\kpc$ to 
$z_h=0.2\kpc$ over a timescale $t_c$, remaining constant thereafter. We assume 
a linear-exponential star-formation history, $\mdotstar(t)\propto te^{-t/\tsf}$.
With a star-formation efficiency timescale $\tau_*=M_g(t)/\mdotstar(t)=2\Gyr$, 
an outflow mass-loading factor $\eta=\dot{M}_{\rm out}(t)/\mdotstar(t)=1.5$, 
$\tsf=3\Gyr$, and $t_c=2.5\Gyr$, the model reproduces the observed locus of 
inner disk stars in $\afe$-$\feh$ and the metallicity distribution functions 
(MDFs) measured by H15 at $|z|=0-0.5\kpc$, $0.5-1\kpc$, and $1-2\kpc$.  
Substantial changes to model parameters lead to disagreement with the H15 data; 
for example, models with $t_c=1\Gyr$ or $\tsf=1\Gyr$ fail to match the observed 
MDF at high-$|z|$ and low-$|z|$, respectively.  The inferred scale-height 
evolution, with $z_h(t)$ dropping on a timescale $t_c \sim \tsf$ at large 
lookback times, favors upside-down formation over dynamical heating of an 
initially thin stellar population as the primary mechanism regulating disk 
thickness. The failure of our short-$t_c$ models suggests that any model in 
which thick disk formation is a discrete event will not reproduce the 
continuous dependence of the MDF on $|z|$ found by H15.  Our scenario for the 
evolution of the inner disk can be tested by future measurements of the 
$|z|$-distribution and the age-metallicity distribution at $R=3-5\kpc$.
\end{abstract}


\section{Introduction} \label{sec:intro} 
Stars in the vicinity of the sun display a double-exponential distribution of
heights $|z|$ above the disk midplane, often described as the 
superposition of a ``thin
disk" and ``thick disk" \citep{gilmore1983,juric2008}. 
Young stars have cool kinematics that confine them to
the thin disk, while stars with hotter kinematics are, on average, older, less
metal rich, and enhanced in the $\alpha$-elements produced by core collapse
supernovae. Abundance studies show that the distribution of [$\alpha$/Fe] is
bimodal at solar or sub-solar [Fe/H], allowing chemical definition 
of ``thin" and ``thick" disk populations that display distinct kinematics 
\citep{bensby2003,lee2011}. A variety of
mechanisms have been proposed for the origin of the thick disk, including debris
of an accreted satellite \citep{abadi2003}, 
heating of a pre-existing disk by satellites
\citep{villalobos2008}, 
gas-rich mergers or a turbulent gas-rich phase in the early star-forming
disk \citep{brook2004,bournaud2009,forbes2012}, 
or radial migration of stars outward from the kinematically hot inner
disk \citep{schoenrich2009b}. 
Two basic questions about the disk are: (1) were thick disk stars
formed in a thick layer or heated over time from a 
thin distribution, and (2) was 
the formation of the thick disk a discrete event or 
extended over many orbital times?

In support of the ``continuous" view, \cite{bovy2012a} 
show that mono-abundance populations in the disk, each defined by a narrow 
bin of $\feh$ and $\afe$,
individually have exponential radial and vertical profiles. 
The superposition of 
multiple populations with different scale-heights 
gives rise to the double-exponential
vertical profile observed for the full 
disk population and accounts for the correlation
between [$\alpha$/Fe] and kinematics. Inspired by this finding, 
\cite{bird2013}
show that similar behavior arises for mono-age populations in a high-resolution
hydrodynamic simulation of the 
formation of a Milky Way-like galaxy, starting from 
cosmological initial conditions. Furthermore, 
\cite{bird2013} show that the
oldest disk stars are born in a vertically thick distribution from a turbulently
supported interstellar medium (ISM), in accord with previous numerical and analytic
arguments about the influence of gravitational instability and feedback on the scale
height of the star-forming gas disk \citep{bournaud2009,forbes2012}. 
They dub this scenario ``upside-down" 
disk formation, by analogy to ``inside-out" growth in which the disk scale length
increases with time. While subsequent heating does increase the scale-height of older
populations (\citealt{bird2013}, fig.~19, and J. Bird et al., in prep.),
cosmological simulations suggest that the distinction between
the thick and thin disks was largely imprinted at birth.

In this paper, we show that the combination of upside-down disk formation with a simple
chemical evolution model can naturally explain the observed [$\alpha$/Fe]-[Fe/H] 
distributions and metallicity distribution functions (MDFs) of the inner disk. The APOGEE
survey \citep{majewski2016} has allowed mapping of the [$\alpha$/Fe]-[Fe/H] distribution over much of
the Milky Way disk through high-resolution near-infrared spectroscopy of more than 100,000
evolved stars. In most regions of the Galaxy, as in the solar neighborhood, stars lie
on two distinct sequences in [$\alpha$/Fe]-[Fe/H], though the locus of high-$\alpha$ stars 
appears to be independent of position within the disk (\citealt{hayden2015}, 
hereafter H15).
However, in the innermost radial zone mapped by H15, Galactocentric radius
$3\kpc<R<5\kpc$, 
stars lie along a sequence that resembles the evolutionary track 
of a simple one-zone chemical evolution model 
(e.g., \citealt{talbot1971,tinsley1980}).
In such a model, low metallicity stars 
occupy a plateau in [$\alpha$/Fe] that reflects the characteristic yields of core collapse supernovae, averaged over the stellar initial mass function (IMF). Once Type Ia supernovae become an important source of enrichment, they further increase [Fe/H], but
they drive down [$\alpha$/Fe] because they produce mainly iron-peak elements. The
location of the ``knee" where [$\alpha$/Fe] turns downward depends mainly on the 
star formation efficiency, while the value of [Fe/H] at late times depends mainly on
the efficiency of galactic outflows, which controls the 
equilibrium between continuing
enrichment from supernovae and loss of metals to stars and winds 
(\citealt{andrews2016},\citealt{weinberg2016}, hereafter WAF).
At $R=3-5$ kpc, most stars far from the midplane 
have enhanced [$\alpha$/Fe] and sub-solar [Fe/H], while most stars near the midplane
have [$\alpha/\mathrm{Fe]}\approx 0$ and supersolar [Fe/H]. We will show that this
distribution can be readily explained by one-zone chemical evolution in a vertically
contracting disk.

Explaining the bimodal [$\alpha$/Fe]-[Fe/H] distribution seen at larger $R$ requires
something more complex than the chemical evolution models considered here, such as
mixing of stellar populations by radial migration 
\citep{sellwood2002,schoenrich2009a}, sharp transitions in
the Galaxy accretion and star formation history \citep{chiappini1997}, 
or an increase of outflow 
efficiency at late times that induces reverse evolution
towards lower [Fe/H] (WAF). Here we focus on the inner disk, 
where simple models
appear applicable. We suspect that the more complex structure of the outer disk
arises because radial migration has a larger impact there,
while the inner disk is dominated by stars that remain close to their birth radius.
Tentative support for this view comes from the modeling of MDFs 
presented by H15 (see their Figure~10). 
A full model for the formation and evolution of the Milky Way
disk must account for its
full spatial and chemo-dynamical structure and explain the 
distributions of elements that trace different nucleosynthetic pathways. This paper
presents idealized models intended to highlight one aspect of this evolution.

\section{Data and sample selection}

Our data sample is a subset of that studied by H15, drawn from the APOGEE survey
\citep{majewski2016}
of the Sloan Digital Sky Survey III (SDSS-III, \citealt{eisenstein2011}). 
It consists of red giant 
and sub-giant stars with elemental abundances inferred from infrared (H-band)
spectroscopy with spectral resolution $\lambda/\Delta\lambda \approx 20,000$ 
\citep{nidever2015}.
Elemental abundances are inferred from the APOGEE Stellar Parameters and 
Chemical Abundances Pipeline (ASPCAP; \citealt{garcia-perez2016}). 
In brief, the parameters of each star
are determined by $\chi^2$ minimization across the full spectrum,
using 6-parameter model grids:
$T_\mathrm{eff}$, $\log g$, an overall metallicity [M/H], 
scalings
of $\alpha$-elements relative to the overall metallicity [$\alpha$/M], and 
individually varied abundances of C and N. 
With free variations of $\alpha$, C, and N,
the metallicity M is responsive mainly to iron peak elements. 
We (like H15) refer to [$\alpha$/M] and [M/H] by the more conventional 
labels [$\alpha$/Fe] and [Fe/H]. 
We apply the same cuts as H15, summarized here in Table
\ref{table:cuts}, to select main sample, evolved stars with high-quality
abundance determinations. These abundances are the ones provided in SDSS DR12 
\citep{alam2015,holtzman2015}.
Further details of APOGEE field and target selection can be found in 
\cite{zasowski2013}.

H15 estimate distances from spectroscopically determined stellar parameters,
PARSEC isochrones \citep{bressan2012}, and extinction-corrected apparent
magnitudes (see H15 and \citealt{holtzman2015} for details).
They use these distances to assign
stars to bins in Galactocentric radius $R$ and vertical distance from the 
midplane $|z|$. 
In this paper we concentrate on H15's innermost radial bin 3 kpc $<R<$ 5 kpc, 
where stars lie along what is plausibly a single evolutionary track. After the 
cuts listed in Table \ref{table:cuts} we are left with 3753 stars: 
2435 at $|z|=0-0.5\kpc$, 849 at $|z|=0.5-1.0\kpc$, and 469 at $|z|=1.0-2.0\kpc$.
The top panels of Figure \ref{fig:hayden_plots} plot 
the metallicity distribution
functions (MDFs) of stars in these three vertical layers, and the bottom panels
plot the distribution of stars in the [$\alpha$/Fe]-[Fe/H] plane, together
with the evolutionary track described in the following section.
Equivalent information in different format appears in Figures~4 and~5
of H15.

\begin{table}[t] 
\begin{center}
{\renewcommand{\arraystretch}{1.2} 
\caption{Cuts made on APOGEE data for sample selection} 
\begin{tabular}{c | c } 
\hline \hline 
Cuts & Notes \\ 
\hline
3 kpc $< R <$ 5 kpc & Inner annulus of H15\\ 
$1.0 < \log g < 3.8$ & Choose giants only\\ 
3500 K $< T_\mathrm{eff} <$ 5500 K & High quality abundances\\ 
S/N $> 80$ & " \\ 
APOGEE\textunderscore TARGET1 bits $\in$ 11,12,13 & Main survey targets only\\ 
ASPCAPFLAG bits $\notin$ 23 & Remove bad stars \\ 
\hline \hline
\end{tabular} } 
\end{center} 
\label{table:cuts} 
\end{table} 

\section{Chemical Evolution Model} \label{sec:models}

We adopt oxygen as the representative $\alpha$-element in
our theoretical model.  The H15 measurements depend on several
$\alpha$-elements with oxygen dominant among them.
Within our model, the defining characteristic
of $\alpha$ elements is that they are produced by
core collapse supernovae only, with a yield that is 
independent of metallicity.

We calculate chemical evolution within a
single galactic zone according to the following specifications:
\begin{enumerate} 

\item The galactic zone consists of a supply of gas out of
which stars are formed over time.  This gas is initially pristine hydrogen and
helium, and it is homogeneously mixed at all times. Any new metals introduced 
by supernovae are immediately mixed into the existing gas supply.  

\item
\label{item:sfe} 
Stars are formed from the gas supply with a constant efficiency 
$\dot M_*/M_g$, which we specify in terms of a star formation 
efficiency (SFE) timescale $\tau_*=M_g/\dot M_*$.  
We choose $\tau_*=2.0$ Gyr, based on the observationally inferred
timescale for molecular gas in local disk galaxies \citep{leroy2008}.

\item The star formation rate (SFR) evolves in a linear-exponential fashion,
i.e.,
\begin{equation} \label{eq:starformation}
\mdotstar(t)=\mdotstar(\tsf) \times \left(\frac{t}{\tsf}\right) e^{-t/\tsf} ~,
\end{equation} 
where $\tsf$ is the star formation timescale.  
Together with assumption~\ref{item:sfe}, this implies a linear-exponential
form for the gas supply $M_g(t)=\tau_*\mdotstar(t)$.
We consider several different values for $\tsf$. 
The Appendix shows results for an exponential star formation
history, but the linear-exponential model corresponds better to the 
predictions of galaxy formation simulations \citep{simha2014},
and it yields a somewhat better fit to the H15 data.

\item Outflow of gas from the zone, due to stellar feedback,
occurs at a rate $\dot M_\mathrm{out}$.  We parameterize outflow efficiency
with the mass-loading factor $\eta$, such that $\eta=\dot M_\mathrm{out}/\dot M_*$.
We adjust the value of $\eta$ depending on $\tsf$ 
to produce similar peaks in the [Fe/H] distribution (see discussion below).

\item Each population of stars forms with masses distributed according to a 
\cite{kroupa2001} IMF, which prescribes a double power-law form for 
\begin{equation} 
\zeta(M) \propto \frac{dN(M)}{dM}
\end{equation}
with slopes $a$ and $b$ above and below $c M_\odot$. We truncate the IMF at
$\Mmin = 0.08M_\odot$ and $\Mmax=100 M_\odot$.

\item There are two sources of new metal production:
core-collapse supernovae (CCSNe) and Type Ia supernovae (SNIa).  
We define the yield parameters $\mocc$, the mass of oxygen produced
by CCSNe per unit mass of star formation, $\mfecc$, the corresponding
quantity for iron, and $\mfeIa$, the mass of iron produced by SNIa
over a time interval 12.5 Gyr per unit mass of star formation.
We adopt $\mocc=0.015$ and $\mfecc=0.0012$, based on a Kroupa IMF
and the CCSN yields of \cite{chieffi2004} and \cite{limongi2006}.
We adopt $\mfeIa=0.0017$ based on the SNIa rate of \cite{maoz2012b} and
an iron yield of $0.77 M_\odot$ \citep{iwamoto1999} per supernova
(see \citealt{andrews2016} for more detailed discussion).
These yields are defined to be net quantities: they do not include the
amount of oxygen and iron incorporated into the stellar population from the
ISM at birth, but  only the newly produced amounts of these elements.

\item CCSNe go off instantaneously upon the birth of a stellar
population; this is a reasonable approximation because we adopt star formation
histories that are smooth on much longer timescales. However, it does assume 
that the ``lock up" of metals in the warm or hot phases of the ISM 
\citep{schoenrich2009a} can be ignored.

\item SNIa follow a power-law delay time distribution (DTD)
after a minimum delay time $t_d$:
\begin{equation} \label{eq:dtd}
  R(t)=
	\begin{cases}
		R_0t^{-1.1} & t>t_d\\ 
		0           & t<t_d ~.
	\end{cases}
\end{equation} 
The form is based on \cite{maoz2012b}, and the normalization $R_0$
cancels out of our calculations once we have specified the
iron yield $\mfeIa$.  We adopt $t_d=0.15\Gyr$.

\item\label{item:recycling} A star of mass $M$ lives for a time
$t_*=10\mathrm{\,Gyr}\times(M/M_\odot)^{-3.5}$, then returns to the
ISM a mass of gas $M-\Mrem$ with the same chemical composition the
star had at birth.  (Stars above $8M_\odot$ additionally return
elements with the CCSN net yields described above.)
The remnant mass as a function of initial mass is
\begin{equation} \label{eq:mrem}
  \Mrem(M)=
	\begin{cases}
		0.394M_\odot + 0.109M & M<8M_\odot \\ 
		1.44M_\odot & M>8M_\odot~.
	\end{cases}
\end{equation} 
(see \citealt{kalirai2008}).  
The lifetime formula can be inverted to yield the main sequence
turnoff mass as a function of time:
\begin{equation} \label{eq:mto}
\Mto(t)=
\left(\frac{t}{10\mathrm{\,Gyr}}\right)^{-1/3.5} M_\odot~.
\end{equation}
\end{enumerate} 

With the assumptions of point~\ref{item:recycling}, the fraction of a
mass of stars formed in a short interval at time $t'$ that has been
recycled to the ISM by time $t$ is
\begin{equation} \label{eq:Frec}
\Frec(t,t')= \frac
  {\int_{\Mto(t-t')}^{\Mmax} M\zeta(M)\left[1-\frem(M)\right]\,dM }
  {\int_{\Mmin}^{\Mmax} M\zeta(M)\,dM }~,
\end{equation}
where $\frem(M) = \Mrem(M)/M$ is the fraction of a star of initial
mass $M$ that is left in the stellar remnant.
While our lifetime formula is inaccurate at high masses, it
adequately captures the important behavior, which is that high
mass stars recycle their envelopes quickly compared to the timescales
for SNIa enrichment and variation of the SFR.  
For a Kroupa IMF, the recycled mass fraction is 
$r = \Frec(t,t') = 0.37$, 0.40, and 0.45 for $t-t' = 1$, 2, 
and 10 Gyr, respectively.  The difference in chemical evolution
between our complete time-dependent recycling and an 
instantaneous approximation with $r=0.4$ is small
(see fig.~7 of WAF).

With these assumptions in place, we may write down the differential equations
governing the amount of oxygen and iron in the ISM.  Metals are injected into
the ISM by CCSNe, SNIa, and recycling, and they are removed from the ISM by
star formation and outflows.  For oxygen,
\begin{equation}\label{eq:ox}
\begin{split} \dot{M}_{\rm O}
  & =  \mocc\mdotstar - \mdotstar\Zo - \eta\mdotstar\Zo \\
  &\hspace{8truemm} + \int_0^t \mdotstar(t')\Zo(t')\Fdotrec(t,t')\,dt' \\
  & = \mocc \frac{M_g}{\tau_*} - (1+\eta)\frac{\Mo}{\tau_*} + 
                  \int_0^t \frac{\Mo(t')}{\tau_*}\Fdotrec(t,t')\,dt' ~,
\end{split}
\end{equation}
where $\Zo = \Mo/M_g$ is the oxygen mass fraction and
$\Fdotrec(t,t')$ is the derivative of equation~(\ref{eq:Frec})
with respect to $t$.
For iron we have the additional contribution of SNIa:
\begin{equation}\label{eq:fe}
\begin{split} \dot{M}_{\rm Fe}
  & =  \mfecc \frac{M_g}{\tau_*} - (1+\eta)\frac{\MFe}{\tau_*} + 
                  \int_0^t \frac{\MFe(t')}{\tau_*}\Fdotrec(t,t')\,dt'\\
  & \hspace{8truemm} + \mfeIa \frac {\int_0^t R(t-t')M_g(t') /\tau_* \,dt'}
                    {\int_0^\infty R(t')\,dt'}~.
\end{split}
\end{equation}

We solve these equations for oxygen and iron evolution with a numerical
code that uses simple Euler integration to evolve the zone from $t=0$ to
$t=12.5$ Gyr, with a step size of $\Delta t=10$ Myr; that is, 
\begin{equation}
M_{\mathrm{X},i} = M_{\mathrm{X},i-1}+\dot M_{\mathrm{X},i-1}\Delta t ~.
\end{equation}
We evaluate the recycling term from the cumulative values of
$\Frec(t,t')$ at the beginning and end of the step. 
We prescribe a star formation history rather than an infall
history (see \citealt{vincenzo2016} for a mathematical discussion
of the relation between these), and the gas mass is therefore
implicitly specified by $M_g(t) = \tau_* \mdotstar(t)$,
with no need to integrate it separately.

Our assumptions are similar to those of many one-zone chemical
evolution models, and they closely follow those adopted for
analytic modeling by WAF.  WAF show that when the integral recycling
terms in equations~(\ref{eq:ox}) and~(\ref{eq:fe}) are replaced
by the instantaneous approximation $r M_X(t)/\tau_*$, then the
resulting equations for oxygen and iron evolution can be solved
analytically if one adopts an exponential DTD for SNIa enrichment
and a constant, exponential, or linear-exponential star formation
history.  These solutions show that for a wide
range of model parameters the values of $\feh$ and $\ofe$ converge
within a few Gyr to ``equilibrium'' values at which sources and sinks
balance, reproducing numerical results \citep{andrews2016}.
The equilibrium $\feh$ depends mainly on yields and on the value of $\eta$,
while the equilibrium $\ofe$ depends mainly on yields.
The importance of equilibrium, and the key role of outflow
efficiency in regulating abundances, has been highlighted in
theoretical studies of the galaxy mass-metallicity relation 
\citep{finlator2008,peeples2011,dave2012}.

In the cosmological simulation of \cite{bird2013}, the median 
height of the star-forming gas layer decreases from about 0.8 kpc
to about 0.2 kpc between $t \approx 2 \Gyr$ and $t \approx 6 \Gyr$,
as the ratio of rotational velocity to turbulent vertical velocity
grows rapidly.  After that, the gas layer continues to get thinner,
but at a much slower rate.  Rather than mimic a single simulation,
we adopt an idealized model in which the scale-height $z_h$
of newly formed stars decreases linearly from 0.8 kpc to 0.2 kpc
over a collapse timescale $t_c$ and remains constant thereafter:
\begin{equation} \label{eq:scaleheight}
  z_h(t,t_c)=
	\begin{cases}
		0.8\kpc-0.6\kpc\times(t/t_c) & t\leq t_c,
		\\ 0.2\kpc & t>t_c ~.
	\end{cases}
\end{equation} 
Figure~\ref{fig:sfr} compares these scale-height histories to the
star formation histories for the values of $t_c$ and $\tsf$ that
we consider below.
In a model where gravitationally driven turbulence of a gas rich
disk governs both the thickness of the gas layer and the efficiency
of star formation \citep{bournaud2009,forbes2012}, we might expect the
collapse time to be linked to the evolution of the SFR.  We do not
impose this condition {\it a priori}, but we find that our fits to 
the H15 data prefer comparable collapse and star-formation timescales.

We evolve our model for 12.5 Gyr.  As noted previously, we choose the
gas supply so that the star formation rate is 
\begin{equation} \label{eq: starformation} 
\frac{dN_*}{dt} \propto \frac{M_g(t)}{\tau_*} \propto t e^{-t/\tsf} ~.
\end{equation} 
The $\Delta N_* = \Delta t \times (dN_*/dt)$ stars formed in a given
timestep are assigned $z$ values
drawn from the distribution 
\begin{equation} 
h(z,t_{c})\propto e^{-z/z_h(t,t_c)} ~,
\end{equation} 
where $z_h(t,t_c)$ is given by equation~(\ref{eq:scaleheight}).  
Thus, for a given $t_c$ the fraction of stars formed at
time $t$ with $z_1<z<z_2$ is given by 
\begin{equation}
\frac{N_{z_1\rightarrow z_2}}{N_\mathrm{tot}}=
  \frac{\int_{z_1}^{z_2}{e^{-z/z_h(t,t_c)}dz}}
       {\int_{0}^{\infty}{e^{-z/z_h(t,t_c)}dz}}~.
\end{equation} 
We ignore any subsequent changes to the vertical structure of the
disk, e.g., from dynamical heating or adiabatic compression, a point
we return to at the end of \S\ref{sec:comparison}.
Each star formed is assigned oxygen and iron abundances
according to the abundances in the ISM at time $t$, which are determined by
the chemical evolution
model described above.  Scatter is added to these values, drawn
from a Gaussian distribution with $\sigma=0.05$ dex.  
We implement scatter partly to enable better visual interpretation 
of our plots and partly to mimic effects of observational errors.

\section{Model Comparison} \label{sec:comparison} 

Figure~\ref{fig:tracks_grid} displays the predicted distribution of stars
in $\ofe-\feh$, in the form of color-coded 2-d historgrams, in the
same three vertical layers used in Figure~\ref{fig:hayden_plots}.
We consider $t_c$ values of 1.0, 2.5, and 4.5 Gyr and 
$\tsf$ values of 1.0, 3.0, and 6.0 Gyr (see Fig.~\ref{fig:sfr}),
making nine models in total.
In each panel, the central track shows the model abundances
calculated without scatter, and colored points mark the abundance
values at $t=0.1$, 0.5, 1.0, 2.0, 5.0, and 10.0 Gyr.
The tracks and colored points are identical in a given column of
the plot because $z$ and $t_c$ do not enter into the chemical evolution.
As explained by WAF, equilibrium abundances depend mainly on
yields and $\eta$ if $\tsf$ is long, but as $\tsf$ approaches
the gas depletion time $\tau_*/(1+\eta-r)$ the equilibrium
abundance grows, and the approach to equilibrium gets slower.
For $\tsf=6.0\Gyr$ and $3.0\Gyr$, we have adopted 
$\eta=1.17$ and 1.50, so that both models yield 
$\feh \approx +0.25$ at $t=10\Gyr$.
For $\tsf=1.0\Gyr$ we choose $\eta=2.83$, which gives $\feh\approx +0.25$
at $t=5\Gyr$, but because of the rapid truncation of the SFR most
stars form below this metallicity.
Tracks and timestamped abundances are fairly similar in the 
$\tsf=6.0$ and $3.0\Gyr$ models, though one can see the slightly 
slower approach to equilibrium for $\tsf=3.0\Gyr$.
The evolutionary track and, especially, abundance vs.\ time
are significantly different for $\tsf=1.0\Gyr$.

As shown previously in Figure~\ref{fig:hayden_plots}, the evolutionary
track of our $\tsf=3.0\Gyr$ model follows the observed locus of H15
stars quite well.  This agreement indicates that, given our adopted
CCSN and SNIa yields, our adopted values of $\tau_*$ and $\eta$
are reasonable.  We note in particular that a much higher (lower)
star formation efficiency would drive the knee of the evolutionary
track to higher (lower) $\feh$, in disagreement with the data,
though we have not investigated the degree to which a change in
$\tau_*$ could be compensated by changes in the star formation
history or SNIa DTD.  The predicted $\ofe$ at high $\feh$ is
slightly sub-solar, while the APOGEE measurements yield solar $\afe$.
Given the uncertainties in observational calibrations, solar abundance
values, and supernova yields, we do not ascribe much significance
to this discrepancy, but models with longer $\tsf$ and thus
higher equilibrium $\ofe$ would fit the data better with other
assumptions held fixed.  Physically, a more rapidly declining 
star formation history leads to lower $\ofe$ at late times because
the ratio of CCSNe to SNIa is lower.

The choices of $t_c$ and $\tsf$ affect the distribution of stars, both along
the tracks and among layers of the disk.  
Examining the middle column isolates the effect of $t_c$.
The most distinctive impact can be seen in the $|z|=1-2\kpc$ layers 
of the $t_c=1.0$ and $2.5\Gyr$ models.
For the shorter collapse timescale,
the distribution of stars in this layer is bimodal, with one
group of stars near the knee of the track
(formation times centered at $t\approx 0.5\Gyr$)
and a second group near the end of the track
(formation times centered at $t\approx 5\Gyr$).
The first group forms during disk collapse, when the scale-height
is still fairly high.  By $t=1.0\Gyr$ the scale-height has dropped
to 0.2 kpc, and although the SFR between $t=1\Gyr$ and $3\Gyr$ is
high, only a modest fraction of all stars are formed in this interval
(see Figure~\ref{fig:sfr}), and a small fraction of these
have $|z| > 1\kpc$.  The disk remains thin at later times, but 
more than 60\% of the stars form between 3 and 10 Gyr, and the
high-$|z|$ tail of this population is sufficient to produce the 
second group of stars, at high $\feh$.
The central model, with $t_c=2.5\Gyr$, still has a bimodal
distribution of stars in the high-$|z|$ later, but now
the $\alpha$-enhanced low metallicity group is far more prominent
because the disk collapse is slower, and more stars form while
the scale-height remains high.

Other differences in the middle column of Figure~\ref{fig:tracks_grid}
are more subtle but follow similar trends.
In the high-$|z|$ layer, the mean metallicity of the
low-$\feh$ group increases as $t_c$ increases, since the
population can evolve toward higher metallicity before the
disk contracts to low scale-height.  
For $t_c=1.0\Gyr$, the mid-$|z|$ layer is
dominated by stars formed at late times when the disk is thin,
but for $t_c=4.5\Gyr$ this layer is dominated by intermediate-age
stars formed during collapse.  The central, $t_c=2.5\Gyr$ model
has the broadest distribution of age and $\feh$ in the mid-$|z|$ 
layer because these two groups join to make a single continuous 
distribution.  The low-$|z|$ layers appear similar in all three
models because a large fraction of stars form after the disk has
reached its 0.2 kpc scale-height.  

Each 2-d histogram in
Figure~\ref{fig:tracks_grid} is normalized relative to the
maximum density in the same layer, similar to the analogous
figure of H15 (their fig.~4).  However, the relative number
of stars in the higher $|z|$ layers depends strongly on $t_c$
and to some degree on $\tsf$; for example, in the central column
the fraction of stars with $|z|=1-2\kpc$ increases
from 1\% to 5\% between the $t_c=1.0$ and $4.5\Gyr$ models.
The relative numbers of stars in the H15 sample are strongly
shaped by the survey strategy and selection function of APOGEE,
so we cannot use them as a model test.  A measurement
of the $|z|$-distribution of stars with $R=3-5\kpc$
provide an additional diagnostic for distinguishing models, a
point we return to in the discussion of Figure~\ref{fig:dndz} below.

The impact of $\tsf$ can be seen by looking across the middle
($t_c=2.5\Gyr$) row of Figure~\ref{fig:tracks_grid}.
A shorter $\tsf=1.0\Gyr$ (left column) leads to a broader
distribution of $\feh$ in the high-$|z|$ layer, since rapid
early star formation allows greater chemical enrichment
before disk contraction.  The low-$|z|$ layer is populated mainly
by stars formed between $t=1$ and $5\Gyr$, but despite the narrower
age range, the metallicity range (roughly $\feh=-0.5$ to $+0.25$) is
broader than that of the corresponding layer for $\tsf=3.0\Gyr$.
Moving to $\tsf=6.0\Gyr$ (right column) leads to a much stronger
concentration of low-$|z|$ stars near the equilibrium metallicity,
both because the approach to equilibrium is faster for longer
$\tsf$ and because this model forms a larger fraction of its stars
at late times (see Figure~\ref{fig:sfr}).  At high $|z|$ this model
shows a bimodal distribution similar to that of the 
($\tsf=3.0\Gyr,t_c=1.0\Gyr$) model discussed previously, again 
with a low-$\feh$ group formed during disk collapse and a
high-$\feh$ group that is the high-$|z|$ tail of the much larger
number of stars formed at late times with a 0.2 kpc scale-height.
Most of the models in Figure~\ref{fig:tracks_grid} show some 
degree of this bimodality at high $|z|$, but the relative importance
of the two groups depends on the specific values of $\tsf$ and $t_c$.
More generally, while there is some tradeoff between these two
parameters, their impact is not strongly degenerate once all three
layers are considered.

Figure~\ref{fig:mdf_grid} plots metallicity distribution functions (MDFs),
the fraction of stars in bins of $\feh$, for the same nine models
and three $|z|$-layers shown in Figure~\ref{fig:tracks_grid}.
Because all stars in a given model lie along a single evolutionary
track, the one-dimensional MDF together with the track itself retains
all of the information from Figure~\ref{fig:tracks_grid}.  The MDF
allows easier comparison to the H15 measurements, shown by the
orange curves in Figure~\ref{fig:mdf_grid}.
The $\tsf=1.0\Gyr$ models are a consistently poor match to the
data in the low-$|z|$ layer, predicting MDFs that are too
broad and too symmetric.  Changes to $\eta$, supernova yields, or
$\tau_*$ could shift the centroid of the MDF,
but the broad and symmetric shape is a generic consequence of the
short star-formation timescale (WAF; \citealt{andrews2016}).
The observed MDF at low $|z|$ instead shows the peaked, negatively
skewed form expected when continuing infall allows a large fraction
of stars to form near the equilibrium abundance, and the $\tsf=3.0\Gyr$
and $\tsf=6.0\Gyr$ models reproduce it much better.
The high-$|z|$ MDF of these models is nearly always bimodal,
but the relative amplitudes of the low-$\feh$ and high-$\feh$ peaks
varies strongly from one $(\tsf,t_c)$ combination to another.
The data do not show a clear bimodality in this layer, and the
$(\tsf=3.0\Gyr,t_c=2.5\Gyr)$ model, for which the two peaks merge
into a continuous broad distribution, provides the best match
to the data.  This model and the $(\tsf=3.0\Gyr,t_c=4.5\Gyr)$ model
also provide the best match to the $|z|=0.5-1\kpc$ MDF,
for which models with shorter $t_c$ or longer $\tsf$ predict too 
many high-$\feh$ stars.  

\begin{table}[t!] 
\begin{center}
{\renewcommand{\arraystretch}{2.2} 
\caption{Statistical quantities used to compare model MDFs to the
H15 data; $x_i$ is the value of [Fe/H] in the $i$th bin, $N$ is
the number of bins, $\Nstari$ is the number of stars in the $i$th bin,
and $\fstari = \Nstari/N_{*,{\rm tot}}$ is the fraction of stars
in the $i$th bin.}
\begin{tabular}{r | l} 
\hline \hline 
Name & Equation \\ 
\hline 
$\Delta^2$ & $\sum_{i=0}^{N}\left(\fstaridat-\fstarimod\right)^2$ \\
Mean & $\bar{x}=\frac{1}{N}\sum_{i=0}^{N}{x_i N_{*i}}$ \\ 
Variance & $\sigma^2=\frac{1}{N}\sum_{i=0}^{N}{(x_i-\bar{x})^2 N_{*i}}$ \\ 
Skewness & $\frac{1}{N}\sum_{i=0}^{N}{\left[(x_i-\bar{x})/\sigma\right]^3 
            N_{*i}}$\\
Kurtosis & $\frac{1}{N}\sum_{i=0}^{N}{\left[(x_i-\bar{x})/\sigma\right]^4
            N_{*i}} - 3$\\ 
\hline \hline 
\end{tabular} } 
\end{center} 
\label{table:stats} 
\end{table}

To quantify the level of agreement seen in Figure~\ref{fig:mdf_grid}, we
compute the squared difference of the MDF histograms
\begin{equation}\label{eq:squarediff}
\Delta^2 = \sum_{i=0}^{N} \left(\fstaridat-\fstarimod\right)^2~,
\end{equation} 
where the sum is over the $N=75$ $\feh$ bins.
Figure~\ref{fig:squarediff} plots these values for the nine
models in each of the three $|z|$ layers.  In the low-$|z|$ layer,
the $\tsf=1.0\Gyr$ model performs very poorly, while the $\tsf=3.0\Gyr$
and $6.0\Gyr$ models reproduce the data about equally well, with
almost no dependence on $t_c$.  In the mid-$|z|$ layer, the
$\tsf=6.0\Gyr$ model fares poorly, especially for low $t_c$, and the
$\tsf=3.0\Gyr$ model performs best at all $t_c$.
The high-$|z|$ layer strongly disfavors $t_c=1\Gyr$, for any
$\tsf$, and it again disfavors the $\tsf=6.0\Gyr$ models at all $t_c$.
Overall, the $\tsf=3.0\Gyr$ models with $t_c=2.5$ or $4.5\Gyr$ are
clearly the most successful, with $t_c=2.5\Gyr$ slightly preferred.
We have not attempted to formally evaluate parameter uncertainties
or a global goodness-of-fit, in part because our models seem too
idealized to warrant such an approach.  In particular, a global $\chi^2$
or maximum likelihood statistic would assign the most weight to the
low-$|z|$ MDF, which is based on the most stars and thus has the
smallest statistical errors, but given that our models are likely
incomplete, the level of agreement in the mid-$|z|$ and high-$|z|$ layers
is a better diagnostic of star formation and disk collapse timescales
than small improvements in reproducing the low-$|z|$ data.
A formal fitting procedure would also require accounting for systematic
uncertainties in the abundance data and for the influence of APOGEE's
evolved star selection on the MDF.

Figure~\ref{fig:statsplots} plots the first four moments of the MDFs ---
mean, variance, skewness, and kurtosis --- for the nine models and 
the data in each $|z|$-layer.  Our precise definitions of these moments
are given in Table~\ref{table:stats}.
In each panel, the horizontal black line marks the measured moment for
the data and the grey band marks the 1-$\sigma$ error on the mean, 
determined from 20 bootstrap resamplings of the original data set.
Because the MDFs are sometimes strongly non-Gaussian (e.g., bimodal),
even four moments do not fully characterize their shapes.
Overall, the moments comparison also shows the $\tsf=3.0\Gyr$ models
with $t_c=2.5\Gyr$ or $4.5\Gyr$ to be the most successful.
However, the $\tsf=6.0\Gyr$ models give better agreement with the
skewness of the low-$|z|$ MDF, and the rejection of $t_c=1.0\Gyr$
models is much more resounding for the $\Delta^2$ statistic than 
it is for MDF moments.

Figure~\ref{fig:tracks_model_data} returns to a more qualitative comparison
between our central model ($\tsf=3.0\Gyr$,$t_c=2.5\Gyr)$ and the
H15 data.  In each layer, we select a number of stars similar to that
in the H15 data, and we plot them (including the 0.05-dex of added
scatter in $\ofe$ and $\feh$) in the same manner as the lower three
panels of Figure~\ref{fig:hayden_plots}, which we repeat here to allow
direct visual comparison.  The model effectively reproduces the
spread and shifting centroid of the metallicity distribution in each layer.
As previously noted, the model predicts a slightly lower $\afe$ at
the highest $\feh$, possibly a symptom of a $\tsf$ that is too low,
or of incorrect supernova yields, or incorrect APOGEE calibration
at super-solar $\feh$.  In the low-$|z|$ layer, the data show a hint
of bimodality in $\afe$ at slightly sub-solar $\feh$, which is not
predicted by the model.  This hint of two distinct sequences could
be a sign that radial migration has moved some low-$\alpha$ stars
from the lower metallicity outer disk into the $R=3-5\kpc$ 
annulus \citep{schoenrich2009a}, the complement to the explanation that
H15 advance for the skew-positive shape of the midplane MDF at
larger radii.

Figure~\ref{fig:dndz} displays the vertical distribution $dN_*/dz$ predicted
by all nine models.  For the central model, $dN_*/dz$ is well
described (up to $|z|=4\kpc$) by the sum of two exponentials with
scale-heights of 0.2 kpc and 0.6 kpc, with equal normalization at
$|z|=1.036\kpc$.  Although our model starts with $z_h=0.8\kpc$, few stars
form at very early times, so the effective scale-height of the 
``thick disk'' component is lower.  We have arbitrarily normalized
all of these distributions to $dN_*/dz=1\kpc^{-1}$ at $z=0$.
With the same mid-plane density, models that have longer $t_c$ at a
given $\tsf$ produce more stars at high $|z|$, as expected.
In the solar neighborhood, the \cite{juric2008} decomposition of
the M-dwarf population implies thin- and thick-disk scale-heights
of 0.3 kpc and 0.9 kpc, respectively, with equal normalizations at
$|z|=0.95\kpc$.  We do not know of any vertical density profile
measurements at $R=3-5\kpc$ where our models apply, and the 
\cite{juric2008} sample is too local for reliable extrapolation
to this Galactocentric radius.  Such a measurement would be a valuable
additional diagnostic for the scenario presented here.
In particular, our choices of $z_h=0.8\kpc$ and 0.2 kpc for the initial
and final scale-heights of our contracting disk, while loosely
motivated by the \cite{bird2013} cosmological simulation,
are somewhat arbitrary.  Additional constraints from a vertical
profile measurement might allow these quantities to be treated
as adjustable model parameters rather than fixed {\it a priori}.

We have framed our model in terms of upside-down disk formation, with
$z_h(t)$ representing the scale-height of the star-forming gas layer
over time.  However, a model in which all stars formed in a thin
layer and dynamical heating puffed up the older stellar populations
would make identical predictions if it produced the same $z_h(t)$.
The principal argument that favors a contracting gas layer over
stellar heating as the predominant mechanism governing
disk thickness is that the $z_h(t)$
of our most successful models is a 
plausible outcome of the upside-down scenario, with disk contraction
occurring as the disk transitions from gas-dominated to 
having comparable gas and stellar mass.
For $\tau_*=2.0\Gyr$, $\tsf=3.0\Gyr$, 
and a recycling fraction $r=0.4$, the ratio
of stellar mass to gas mass is 0.5 at $t=2.5\Gyr$ and
1.2 at $t=4.5\Gyr$.  In a heating scenario, these $z_h(t)$
histories would require a heating mechanism that is sharply
peaked at early times (see Figure~\ref{fig:sfr}), 
rather than the continuous, roughly
power-law dependence on lookback time that is predicted by
typical dynamical heating mechanisms \citep{hanninen2002}.
However, we have not explicitly demonstrated that a
conventional heating model is inconsistent with the H15 data.
Even in an upside-down scenario, dynamical heating is likely
influence the scale-height of populations at the present day
(\citealt{bird2013}, Fig.~19; J.\ Bird et al., in prep.).

We have not explicitly examined scenarios in which disk
heating is a sudden event, as might result from a satellite merger.
Given the drastic failure of our $t_c=1.0\Gyr$ models, however, we think
it unlikely that a discrete heating scenario can reproduce the
steady trend of the H15 abundance distributions with $|z|$.

\section{Conclusions} \label{sec:conclusions} 

Motivated by an ``upside-down'' scenario for thick disk formation,
in which the height of the star-forming gas layer contracts as the
stellar mass of the disk grows
\citep{bournaud2009,forbes2012,bird2013},
we have presented a simple model to explain the $|z|$-dependent $\afe$-$\feh$
distributions of inner disk stars ($3\kpc < R < 5\kpc$) measured
by H15 from the APOGEE survey.  
The H15 inner disk stars appear to lie
along a single evolutionary track, with most high-$|z|$ stars
at sub-solar $\feh$ and $\afe = 0.2-0.4$, most low-$|z|$ stars
at super-solar $\feh$ and solar $\afe$, and mid-$|z|$ stars 
spread broadly over the $\feh$ range $-0.5$ to $+0.25$.  
Our model combines one-zone chemical
evolution with a disk that contracts linearly from a scale-height
of $z_h=0.8\kpc$ to $z_h=0.2\kpc$ over a timescale $t_c$, with
$z_h=0.2\kpc$ thereafter.  
For specified assumptions about 
supernova yields and the SNIa delay time distribution, the
shape of the model evolutionary track is governed mainly by the
outflow mass loading efficiency $\eta$, which sets the end-point
in $\feh$, and the star formation efficiency timescale $\tau_*$,
which sets the location of the knee where the track bends toward
low $\afe$.  The timescale $\tsf$ of our linear-exponential star
formation history, $\mdotstar(t) \propto t e^{-t/\tsf}$, has a 
small impact on the shape of the evolutionary track but a strong
influence on the metallicity distribution function.
Together $\tsf$ and the disk collapse timescale $t_c$ determine the
shape of the MDF in each $|z|$-layer.  Our primary comarison
of models to the H15 data appears in Figure~\ref{fig:mdf_grid}.

The evolutionary track of the H15 sample is well matched by a model
with $\tau_* = 2.0\Gyr$, $\eta=1.5$, and $\tsf=3.0\Gyr$.  With
$t_c=2.5\Gyr$, this model reproduces the observed MDF in each 
$|z|$-layer, and a model with $t_c = 4.5\Gyr$ is almost equally
successful.  With collapse timescale 
$t_c=1.0\Gyr$, however, the model predicts a strongly
bimodal MDF at high $|z|$, contrary to observations.
Models with sharply peaked early star formation, $\tsf=1.0\Gyr$,
predict a low-$|z|$ MDF much broader and more symmetric than
the H15 measurement.  Models with $\tsf=6.0\Gyr$ yield good agreement
with the low-$|z|$ MDF, but for any choice of $t_c$ they fail to
match the observed MDF in the mid-$|z|$ and/or high-$|z|$ layer.
Within our framework, therefore, the H15 data suffice to provide
separate constraints on $\tau_*$, $\eta$, $\tsf$, and $t_c$,
at least at the factor-of-two level.

Our inferred $z_h(t)$ could in principle be explained by either
a contracting star-forming gas layer or age-dependent heating of
stellar populations formed in a thin layer.  A pure heating model
seems unlikely to predict a $z_h(t)$ that rises so sharply at large
lookback times; for example, the scale-height of our $t_c=2.5\Gyr$
model triples between lookback times of $10\Gyr$ and $11.5\Gyr$
(or $8\Gyr$ and $11\Gyr$ for $t_c=4.5\Gyr$).  Conversely, the inferred
$t_c$ corresponds to the interval over which the disk goes from being
gas dominated to having comparable gas and stellar mass fractions,
a natural timescale for disk contraction in the upside-down scenario.
The failure of our short-$t_c$ models makes it unlikely that any
scenario in which thick disk heating is a single discrete event
can explain the continuous $|z|$-dependence of the MDF observed by H15.
However, our model for $z_h(t)$ is obviously idealized, and predictions
of alternative models should be tested directly against the data.
We do not claim that our model for the inner disk is unique,
but it successfully describes the main characteristics of the H15
measurements with simple, physically plausible assumptions and a
small number of adjustable parameters.  

There are a number of directions for future progress in constraining
and testing the scenario presented here.  As illustrated in
Figure~\ref{fig:dndz}, a direct measurement of $dN_*/dz$ for the
inner disk ($R \approx 4\kpc$) would provide strong additional
constraints, and perhaps allow the initial and final disk scale
heights to be treated as adjustable parameters without losing the
power to constrain $t_c$ and $\tsf$.  The DR13 APOGEE data set 
incorporates many improvements in abundance and stellar 
parameter determinations (SDSS Collaboration, 2016), and the APOGEE-2 survey
of SDSS-IV (M. Blanton et al., in preparation) will eventually yield
a substantially larger data set for studying the inner galaxy.
More predictive semi-analytic \citep{forbes2012,forbes2014} or simulation-based
(\citealt{bird2013}; J.\ Bird et al., in preparation) models for
the contracting gas layer could be combined with stellar heating
prescriptions to produce more realistic $z_h(t)$ templates and perhaps
yield informative constraints on the physical parameters of such
models.  Full forward modeling would allow more rigorous statistical
determination of parameter values and uncertainties, accounting for measurement
errors and selection biases, and including nuisance parameters
to represent uncertainties in supernova yields.

\begin{figure}[h]
\centering 
\includegraphics[scale=0.45]{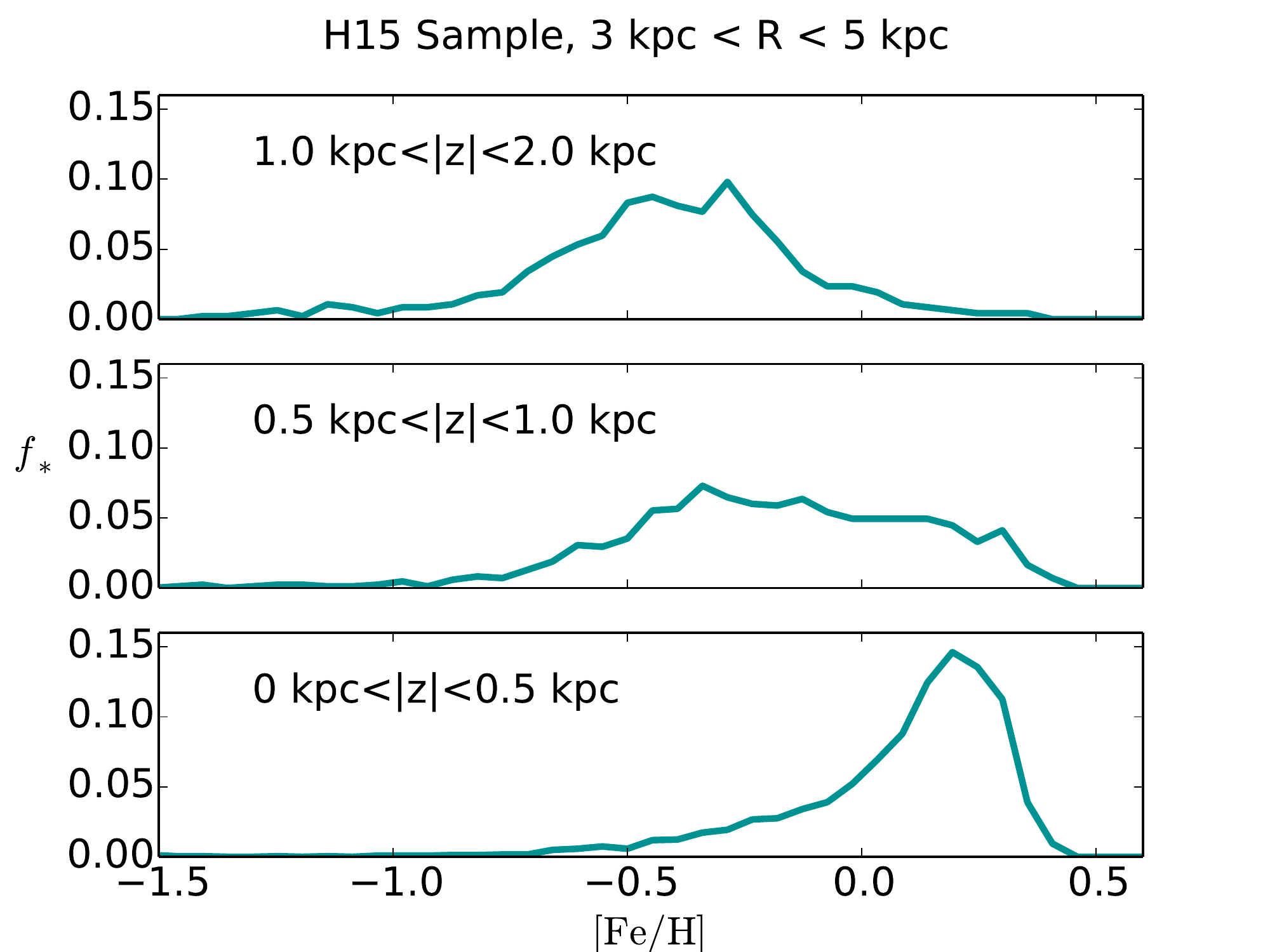}\\
\includegraphics[scale=0.45]{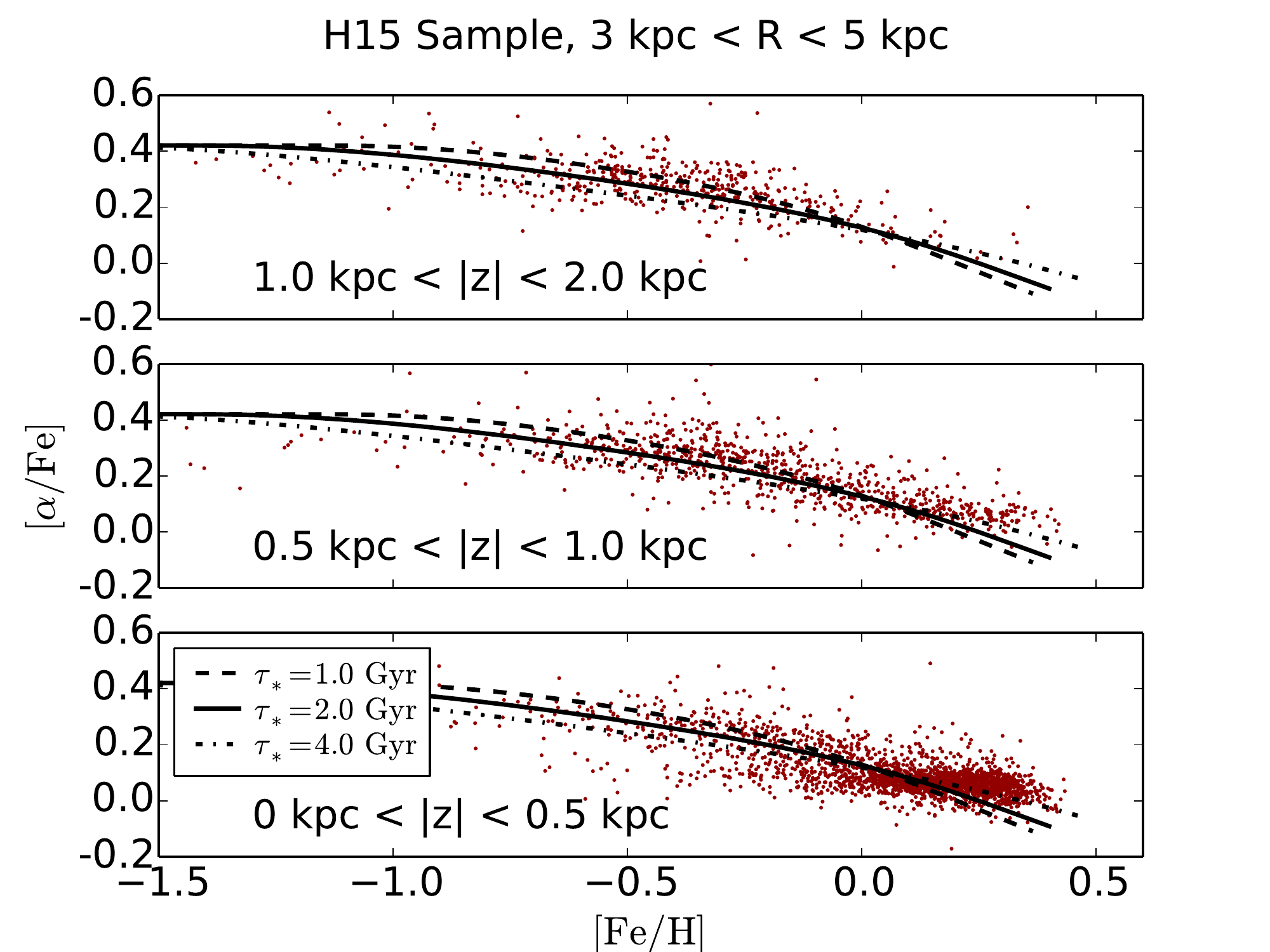}
\caption{(Top) MDFs of stars from the $3\kpc<R<5\kpc$ radial bin of H15, 
in three bins of $|z|$. [Fe/H] bins are	of width 0.053 dex 
(75 bins over the range $-2.5< \feh <1.5$).  
In each panel, the MDF is normalized to the 
total number of stars in that layer.
(Bottom) Distribution of the sample in $\afe-\feh$ space,
with each point representing one star.  The black lines, which are the 
same in each panel, show the evolutionary track predicted by our chemical
evolution code for linear-exponential star-formation timescale 
$\tsf=3.0\Gyr$, outflow mass loading $\eta=1.5$, and star formation
efficiency timescales $\tau_*=1.0\Gyr$, $2.0\Gyr$, and $4.0\Gyr$.
}
\label{fig:hayden_plots}
\end{figure}

\begin{figure}[h]
\centering
\includegraphics[scale=0.6]{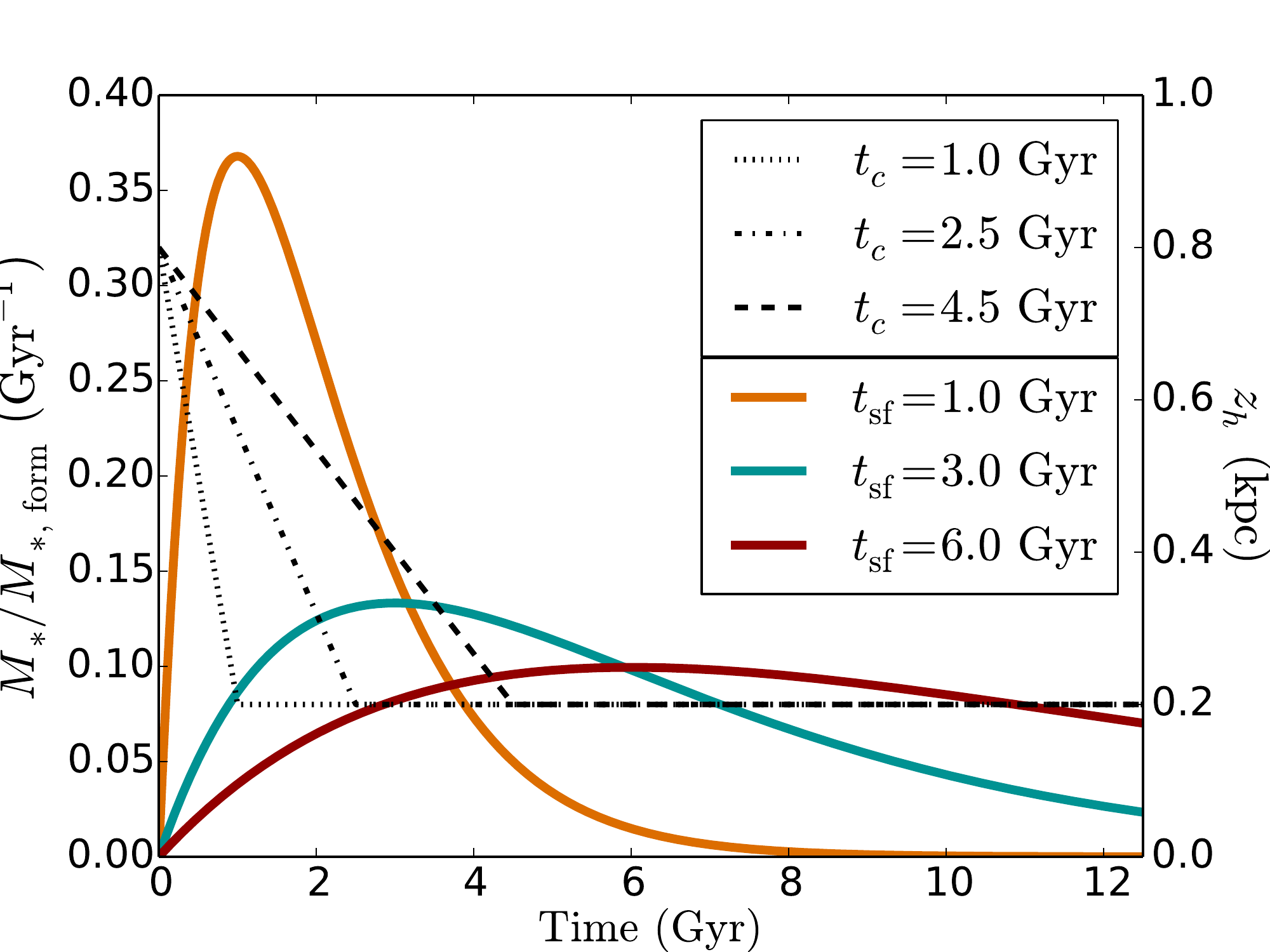}
\caption{
Star formation rate and disk scale height as a function of time
for our models.  The colored curves show the fraction 
of stars formed in each timestep, as indicated by the left axis. 
These are normalized such that the area under each curve is 1.0, 
i.e. the cumulative number of stars is the same in each case after 
12.5 Gyr has passed. The black lines indicate the scale-height 
of the disk over time and correspond to the right axis.}
\label{fig:sfr}
\end{figure}

\begin{figure*}[h]
\centering
\makebox[\textwidth][c]{\includegraphics[width=1.1\textwidth]{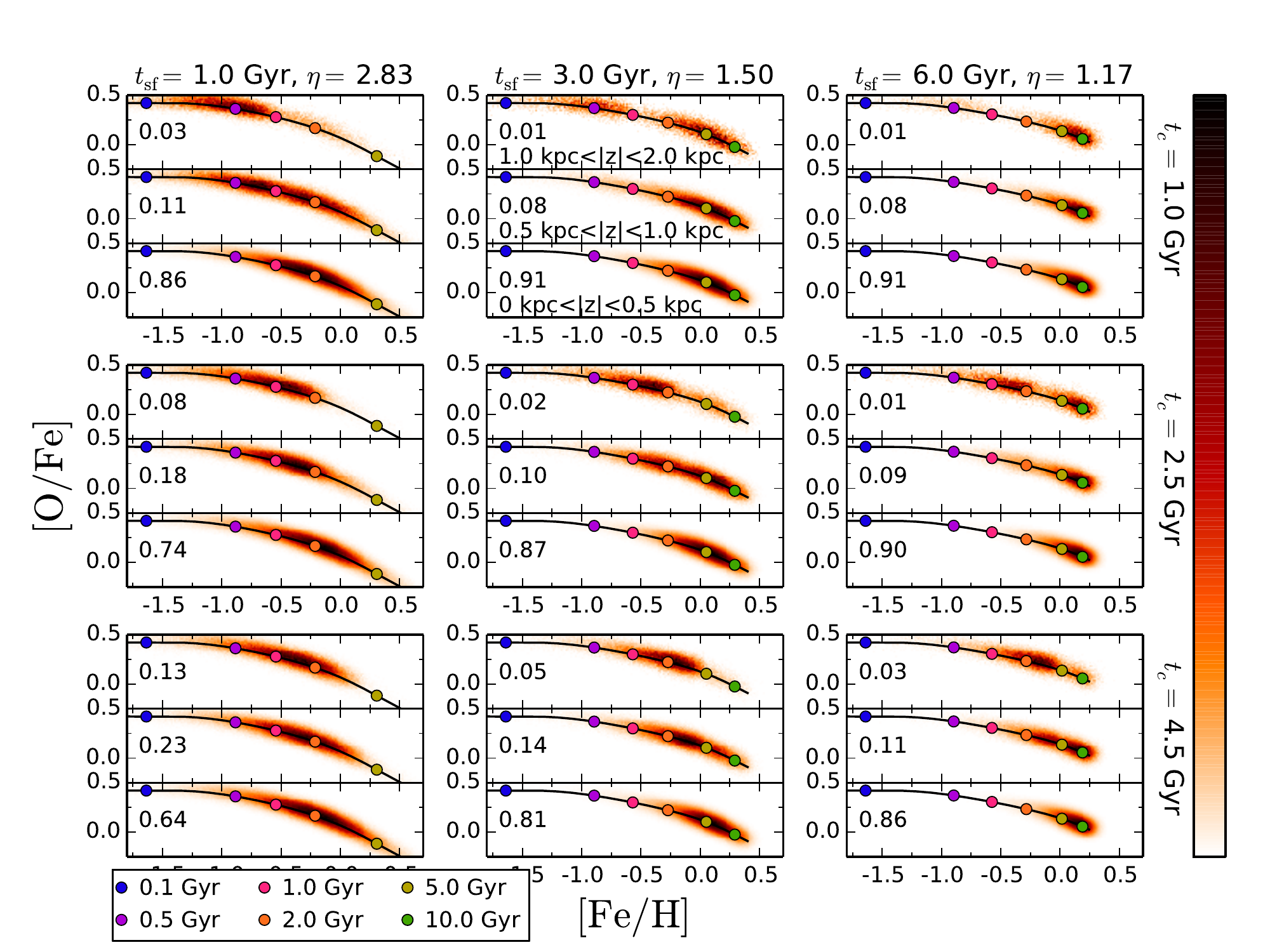}}%
\caption{Predictions of our nine models, plotted as 2D histograms in [O/Fe]-[Fe/H] space. Panels increase in $\tsf$ from left to right and increase in $t_c$ from top to bottom. Each colored circle represents a timestamp corresponding to the times in the legend. Approximately $10^6$ stars are generated between 0.0 and 2.0 kpc in each model instance (the number is not exactly $10^6$ since a small number of stars form above 2.0 kpc); the fraction of these stars in each $|z|$-range are indicated on the lefthand side of each layer. There are 200 bins in the 
$x$-direction and 50 in the $y$-direction. The coloring is normalized such that white indicates 0 stars in a bin and black corresponds to the maximum number of stars in a bin within each layer, with a linear color scale shown on the right; thus, in the lower layers, where there are more stars, the color scale will cover a wider range of star counts than in the upper layers.}
\label{fig:tracks_grid}
\end{figure*}

\begin{figure*}[h]
\centering
\makebox[\textwidth][c]{\includegraphics[width=1.1\textwidth]{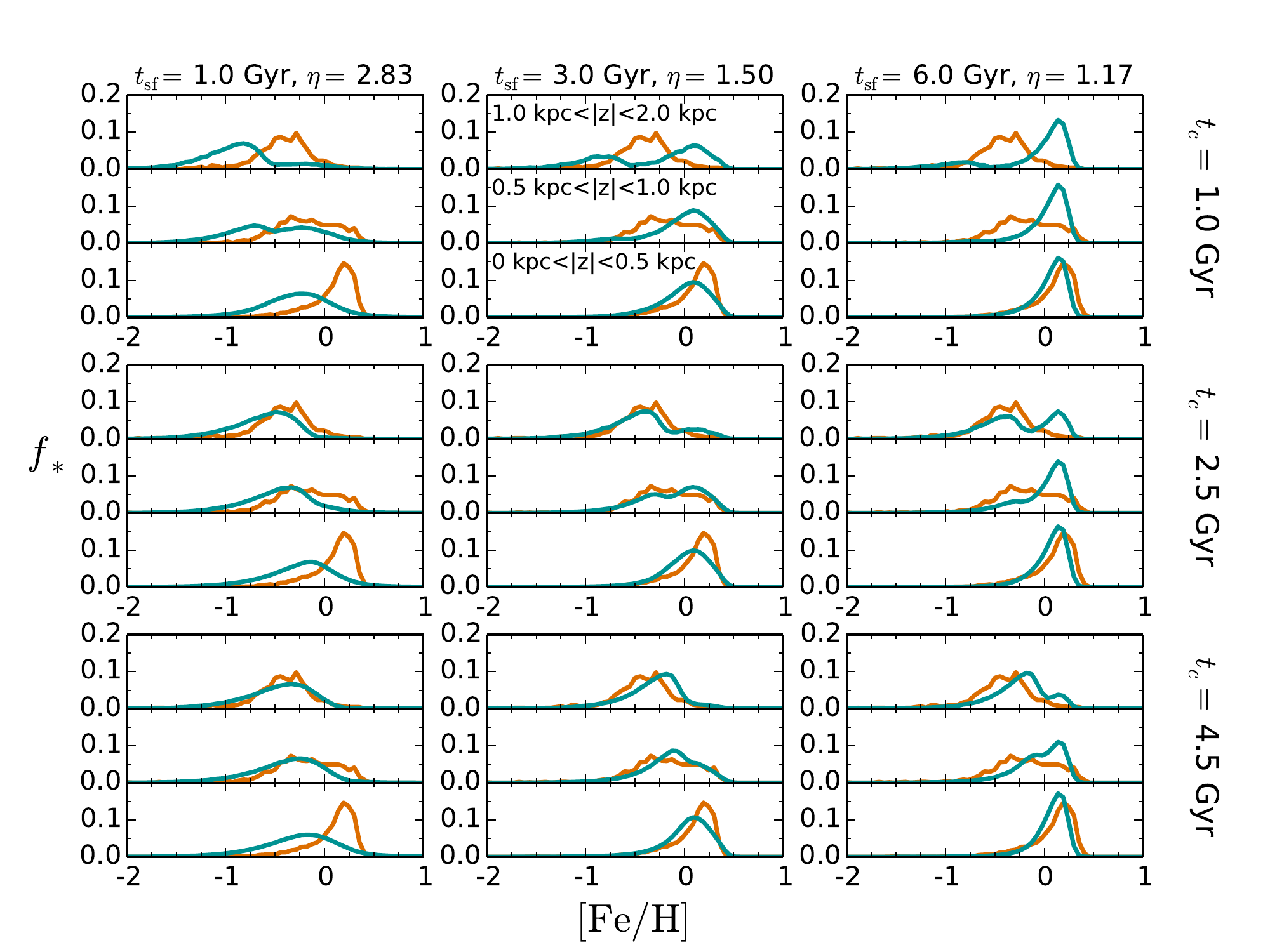}}%
\caption{MDFs in three $|z|$-layers for
our nine models. The MDFs of the H15 stellar sample are plotted in orange.
The blue curves show the model predictions. The panels increase in $\tsf$ from left to right and increase in $t_c$ from top to bottom. MDFs are normalized 
separately in each layer, so that $f_*$ integrates to one in each sub-panel.
}
\label{fig:mdf_grid}
\end{figure*}

\begin{figure}[h]
	\centering
	\includegraphics[scale=0.45]{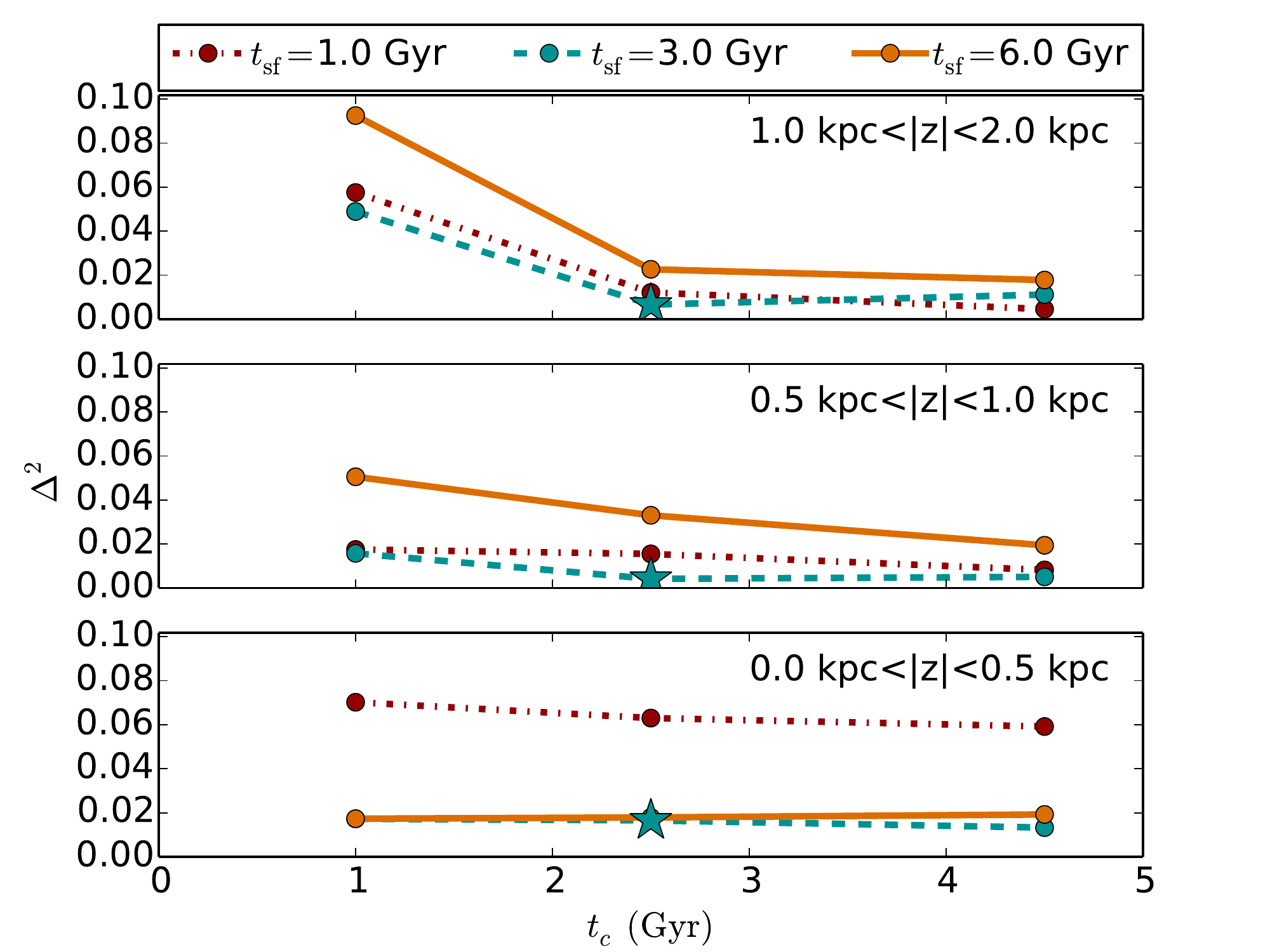}
	\caption{The square difference (eq.~\ref{eq:squarediff}) for the MDFs produced by each of our nine models in each $|z|$-layer. Lines connect models with the same value of $\tsf$, while $t_c$ is indicated on the horizontal axis. The central model is marked with a star in each case.}
	\label{fig:squarediff}
\end{figure}

\begin{figure*}[h]
\centering
\includegraphics[scale=0.4]{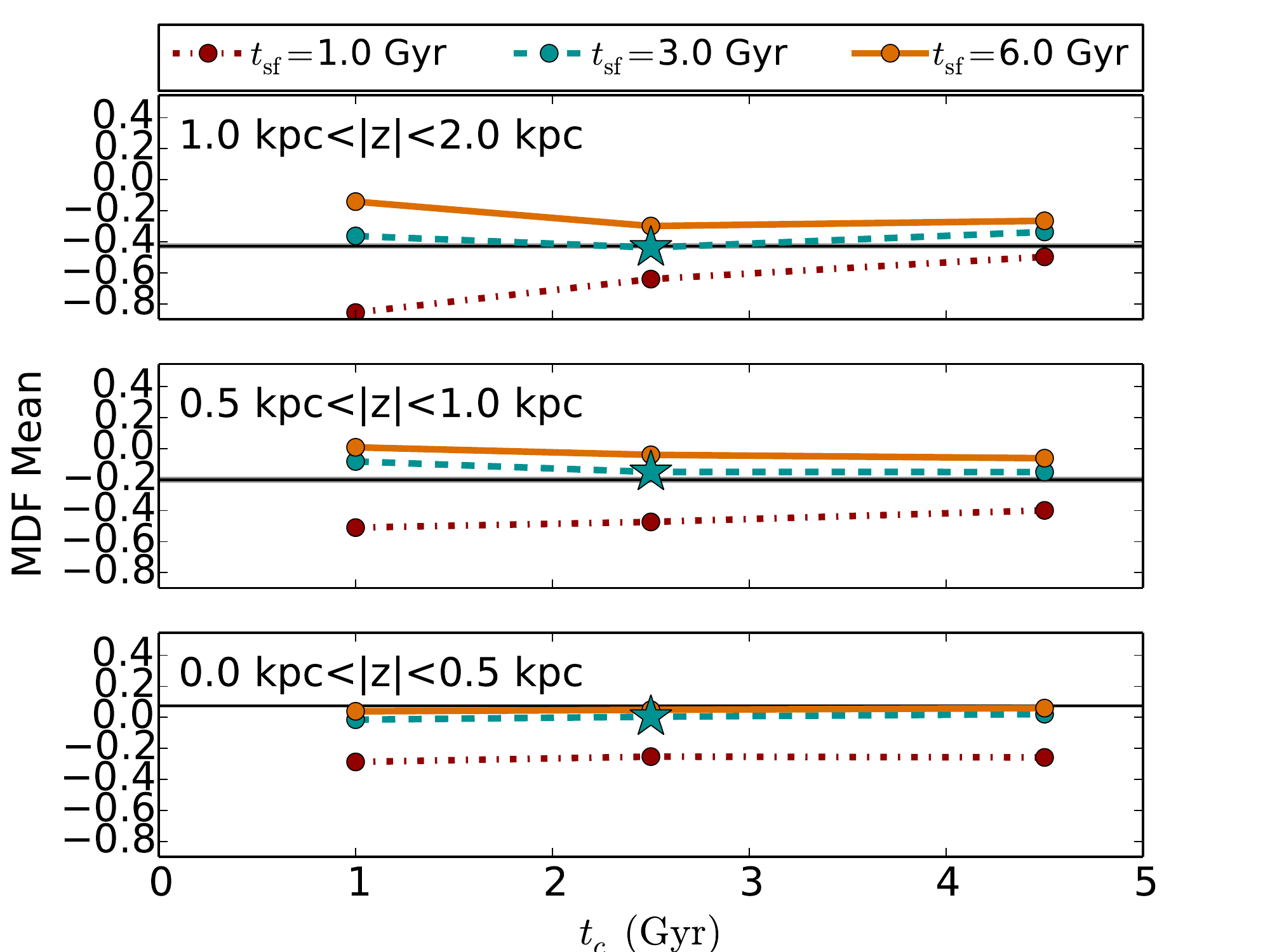}
\includegraphics[scale=0.4]{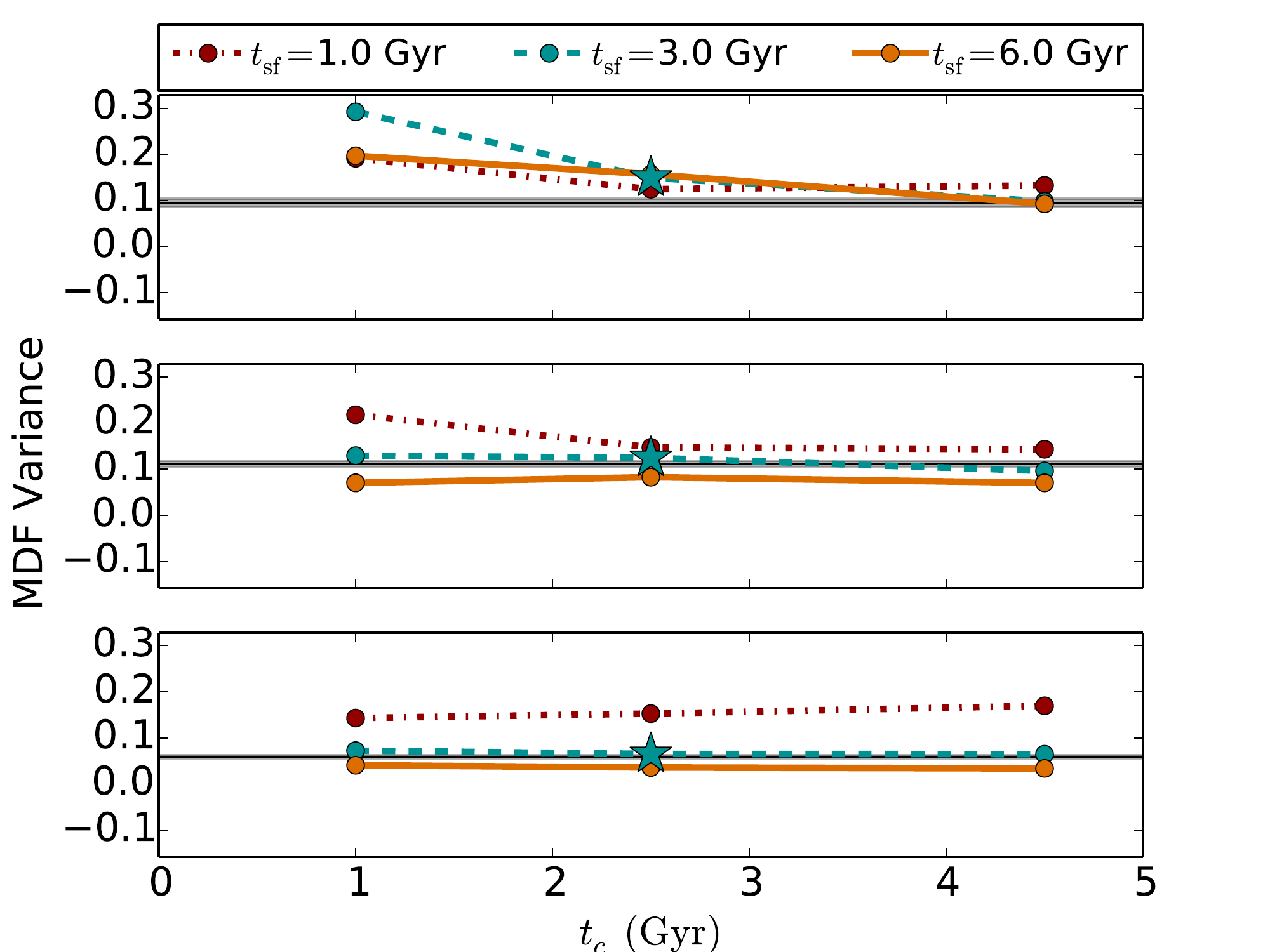}\\
	
\includegraphics[scale=0.4]{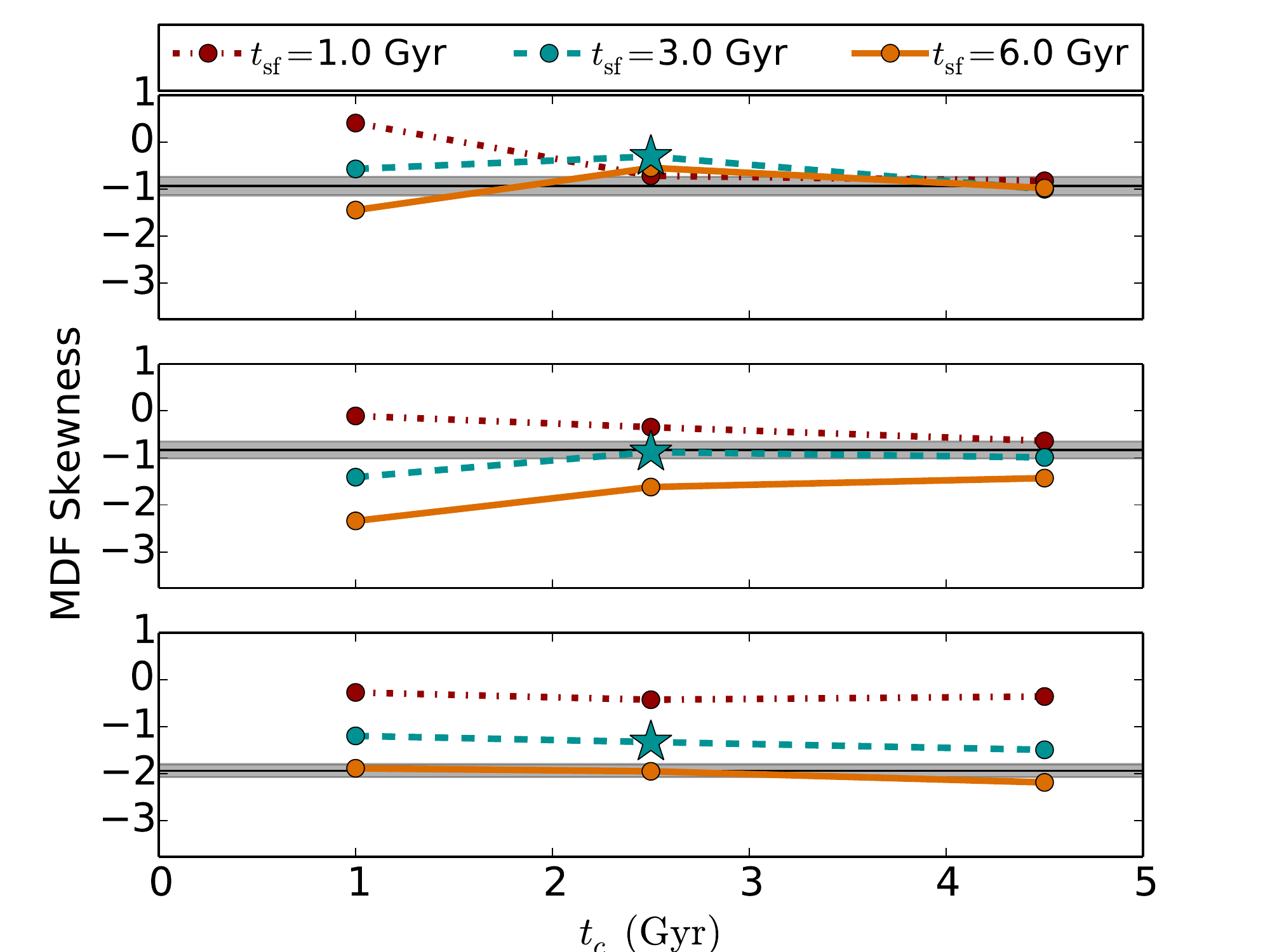}
\includegraphics[scale=0.4]{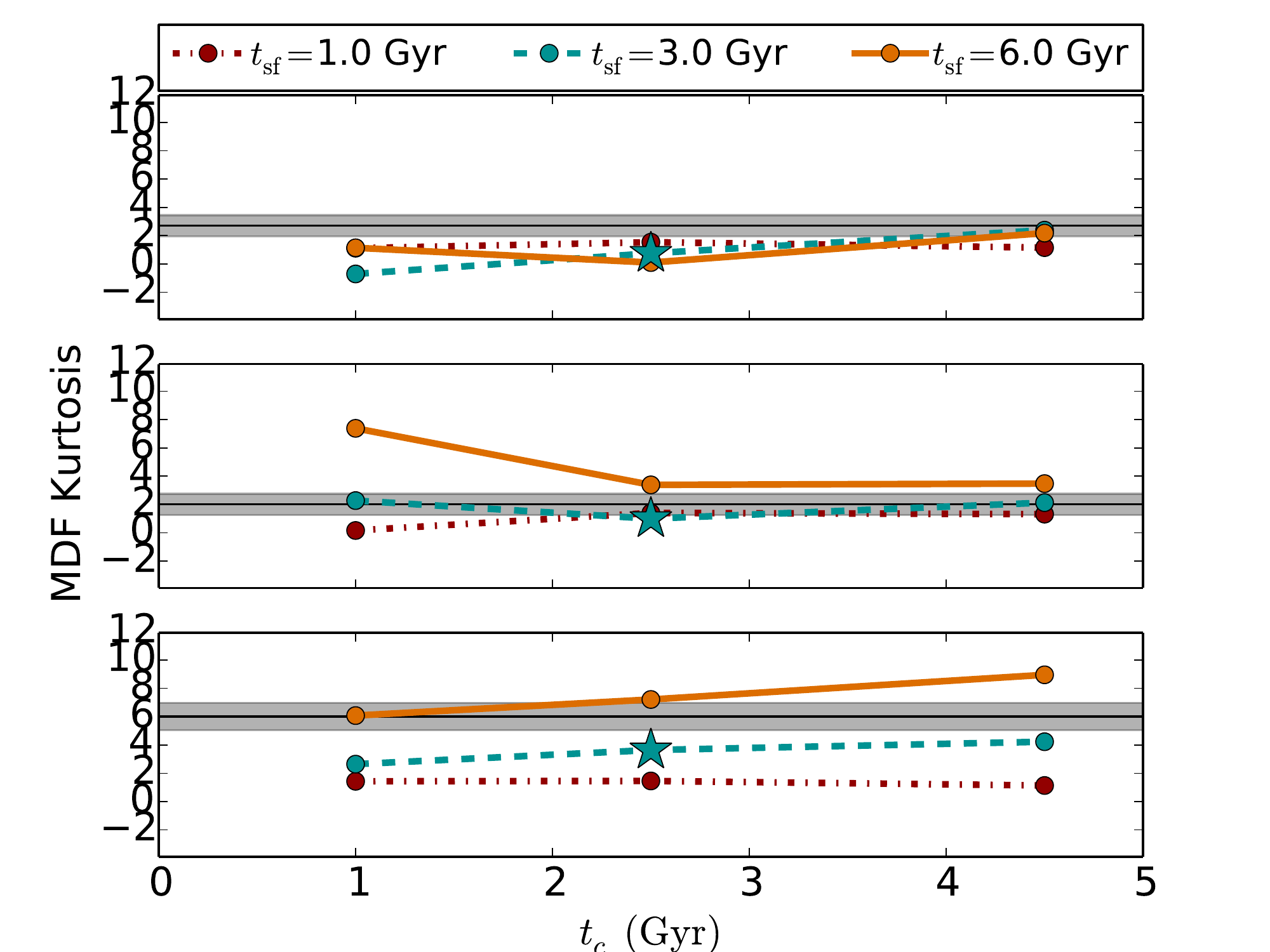}
\caption{Four statistical moments -- mean, variance, skewness, and kurtosis (see Table \ref{table:stats}) -- for the MDFs produced by our nine models in each 
$|z|$-layer, and for the observed MDFs. Lines connect models with the same value of $\tsf$, while $t_c$ is indicated on the horizontal axis. The central model is marked with a star in each case. The horizontal black line is the mean value of each statistic for the data, 
and the gray band denotes the 1-$\sigma$ error on this mean value
as determined by bootstrap resampling of the dataset. 
}
\label{fig:statsplots}
\end{figure*}

\begin{figure*}[h]
\centering
\includegraphics[scale=0.45]{hayden_track_plots.pdf}
\includegraphics[scale=0.45]{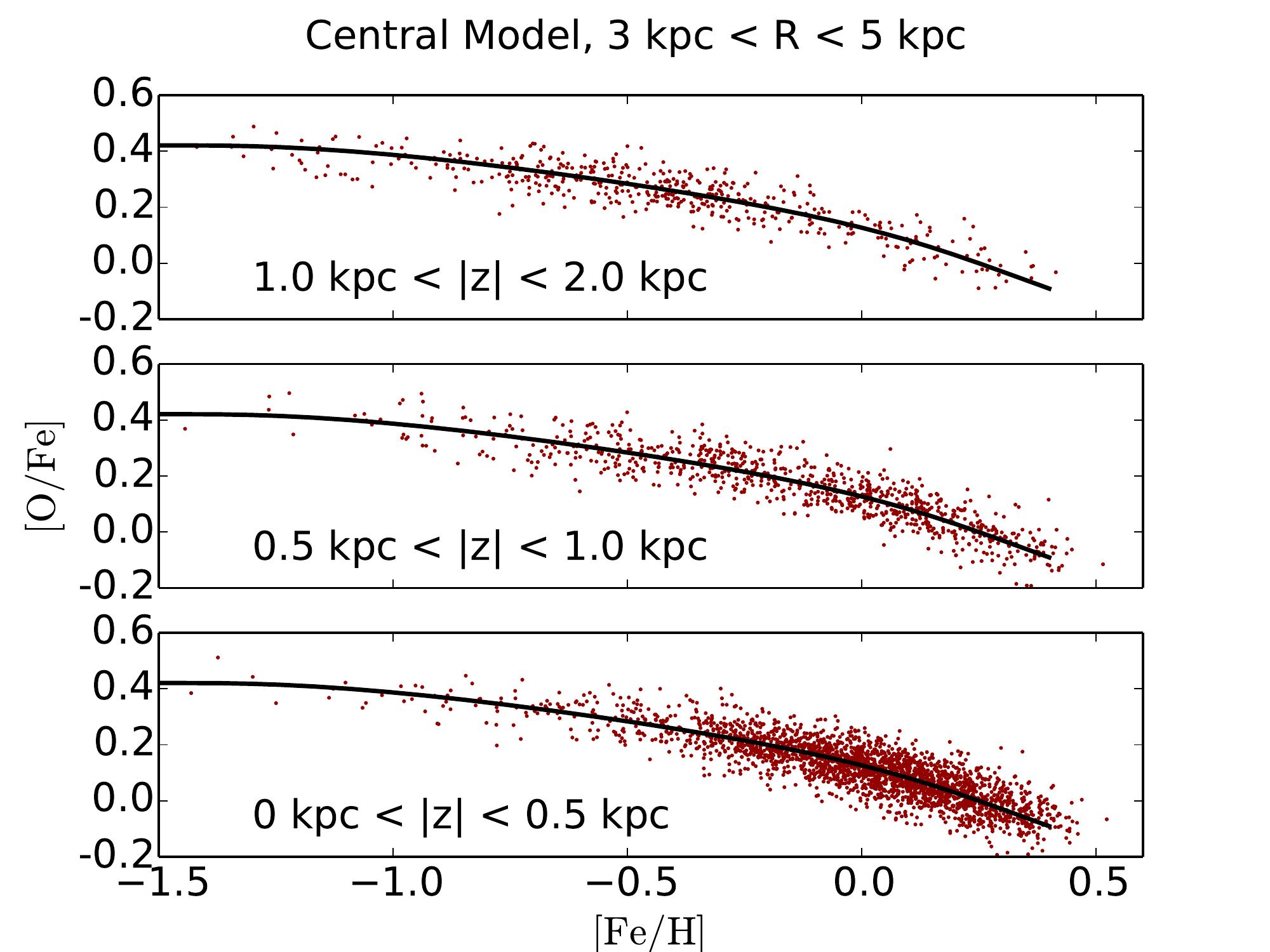}
\caption{The distribution of the H15 stellar sample in 
$\afe$-$\feh$ (top) compared to the predictions
of our central model ($\tsf=3.0\Gyr$, $t_c=2.5\Gyr$)
on bottom. The stars generated by our model have been resampled to reflect the sampling of the data, such that the total number of stars in any given layer are approximately equal.  Scatter of 0.05 dex in $\ofe$ and $\feh$ is added
to the model abundances.
}
\label{fig:tracks_model_data}
\end{figure*}

\begin{figure*}[h]
\centering
\makebox[\textwidth][c]{\includegraphics[width=1.1\textwidth]{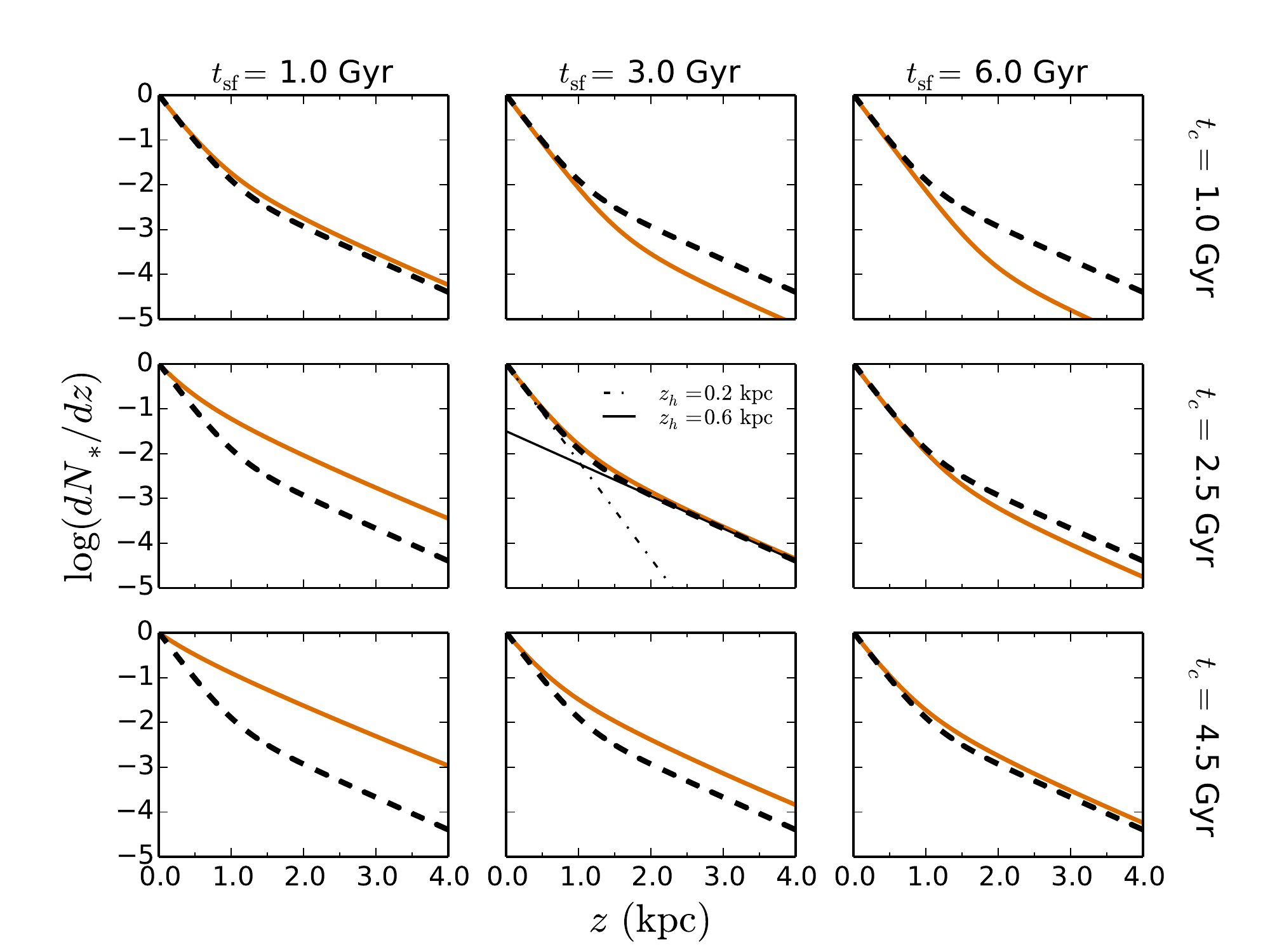}}%
\caption{Vertical distribution $dN_*/dz$ after 
12.5 Gyr, normalized such that $\log(dN_*/dz)=0$ at $z=0$. 
Orange curves show the model predictions.
Black dashed curves, the same in all panels for reference,
show the sum of two exponentials with scale heights of 
$z_h=0.2\kpc$ and $0.6\kpc$, a decomposition that fits
the predictions of our central model.
}
\label{fig:dndz}
\end{figure*}

Stellar ages, inferred from asteroseismology
\citep{pinsonneault2014,stello2015} or chemical abundance signatures
\citep{martig2016,ness2016}, would enable an especially direct test of our
scenario.  In the solar neighborhood, the age-metallicity relation
exhibits a large dispersion \citep{edvardsson1993}, a property
naturally explained by models in which radial migration mixes
stars formed at different Galactocentric radii
\citep{wielen1996,schoenrich2009a}.  We have focused on the
inner disk because, in contrast to the solar neighborhood,
its observed $\afe$-$\feh$ distribution resembles that predicted
by a one-zone chemical evolution model with constant parameters.
If our scenario is correct, then the age-metallicity relation of
the inner disk should have low scatter and no dependence on $|z|$,
and it should lie close to the relation predicted by our $\tsf=3.0\Gyr$
models in Figure~\ref{fig:tracks_grid}.  While radial mixing by
spiral arm perturbations is a symmetric process \citep{sellwood2002},
it may have much less impact on the inner disk because there are
more stars available to migrate from small $R$ to large $R$ than
{\it vice versa} (see H15, figure~10).
The horizontal ridge of stars visible in the low-$|z|$ $\afe$-$\feh$
distribution of H15 (bottom panel of Figure~\ref{fig:hayden_plots}),
while only a hint of the bimodality in $\afe$ that becomes obvious 
at larger $R$, could be an indication that some of these stars
have migrated inward from more distant birth sites.
A complete test of our inner disk scenario should include a realistic
allowance for radial migration from the outer disk.

As emphasized by \cite{nidever2014} and H15, the ``high-$\alpha$ locus''
of stars in the $\afe$-$\feh$ plane is remarkably universal, with
no detected dependence on $R$ or $|z|$.  This universality
could indicate that our evolutionary scenario for the
inner disk describes the Galaxy' early history over a much larger range
of $R$, or that the high-$\alpha$ population throughout the disk
is dominated by stars born at small $R$ and scattered outward by
radial migration.  Assessing these possibilities requires embedding
our model in a more complete framework that can account for the 
bimodality of the $\afe$-$\feh$ distribution observed over much
of the Galactic disk.  As surveys such as APOGEE,
Gaia-ESO \citep{gilmore2012}, LAMOST \citep{luo2015},
and GALAH \citep{desilva2015} extend multi-element maps from the
solar neighborhood to all regions of the Galaxy, the prospects
for constructing, constraining, and testing comprehensive scenarios
for the chemical evolution of the Milky Way grow dramatically brighter.

\acknowledgments 
We thank Jennifer Johnson, Brett Andrews, Jonathan Bird, and 
Michael Hayden for valuable input at stages of this work.
We thank Michael Hayden and Jon Holtzman for providing the
distances used to define our observational comparison sample,
and we acknowledge the immense efforts of the entire APOGEE
team needed to produce this data set.
This work was supported by NSF grant AST-60033987.

\appendix For our main analysis, we chose to adopt a linear-exponential form
for the star formation (eq.~\ref{eq:starformation}).
An alternative and somewhat more conventional choice would be an 
exponential form, i.e.,
\begin{equation} 
\label{eq:starformation2} 
\mdotstar(t) \propto e^{-t/\tsf}~.
\end{equation} 
We consider a linear-exponential model to be more physically motivated
because it does not require starting with a large gas supply already
in place at $t=0$, and because it corresponds better to the star formation
histories predicted in cosmological simulations \citep{simha2014}.
However, it is also interesting to see how our models perform
if we adopt an exponential $\mdotstar(t)$.

\begin{figure*}[h]
\centering
\makebox[\textwidth][c]{\includegraphics[width=1.1\textwidth]{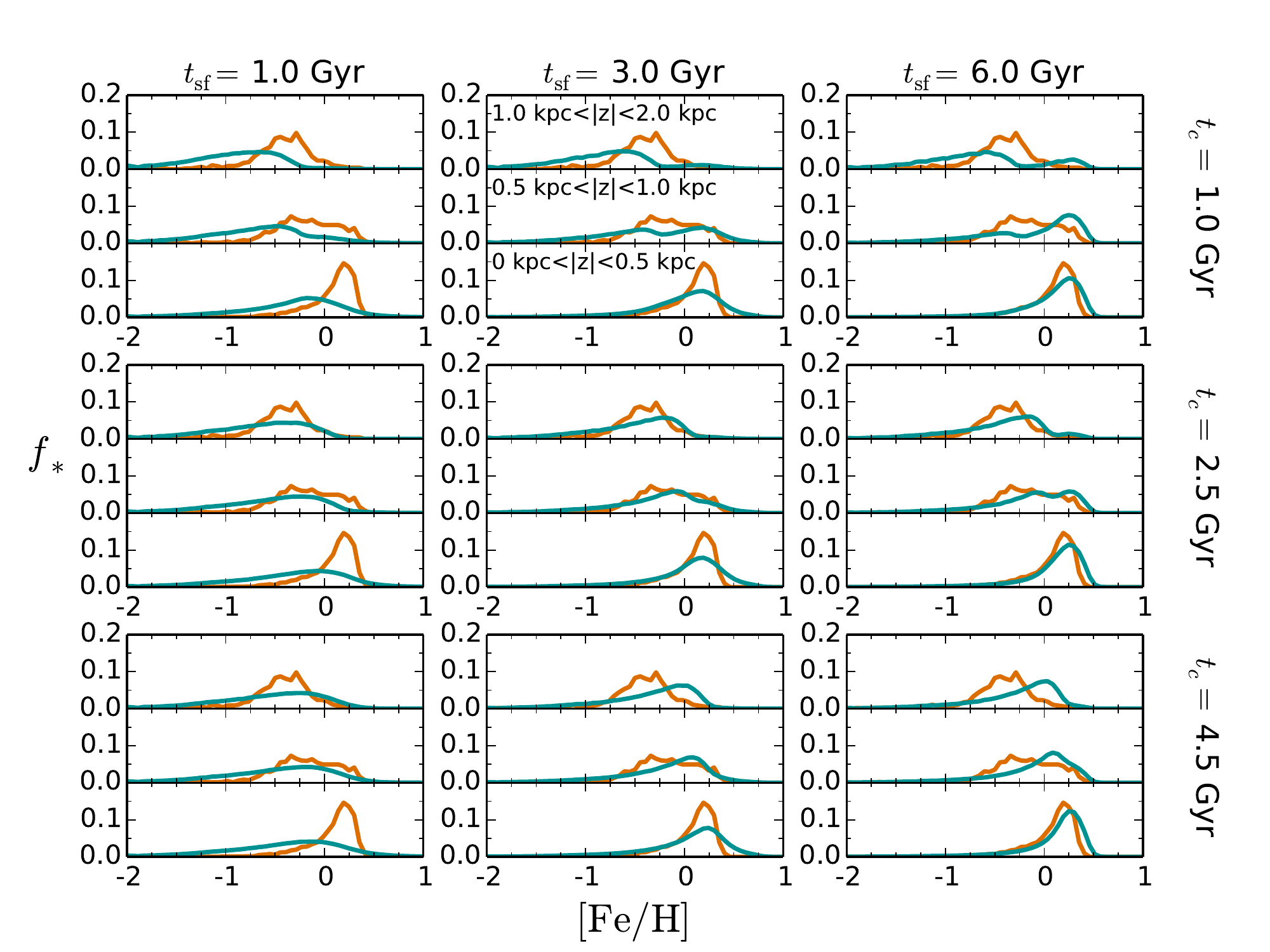}}%
\caption{MDFs for our nine models, in the case where star formation has an exponential form. Orange curves show the observed MDFs and blue curves show the
model predictions.
See the caption of Figure \ref{fig:mdf_grid} for further explanation.}
\label{fig:mdf_grid_exp}
\end{figure*}

Figure~\ref{fig:mdf_grid_exp} repeats the comparison of
Figure~\ref{fig:mdf_grid} for models with an exponential star
formation history.  Most of the trends seen in Figure~\ref{fig:mdf_grid} 
are repeated here.  However, the low-$|z|$ MDF of the $\tsf=3.0\Gyr$
exponential model fits the H15 data much less well than that of the 
corresponding linear-exponential model.  The $\tsf=6.0\Gyr$ model
fares better, and with all layers considered the parameter combination
$(\tsf,t_c)=(6.0\Gyr,2.5\Gyr)$ is the most successful of the
exponential models.  It is somewhat less successful than our
central linear-exponential model at reproducing the shapes of
the mid-$|z|$ and high-$|z|$ MDFs.  We conclude that our modeling
provides moderate but not strong evidence for an inner disk star formation
history that grows over time before turning downwards, rather
than starting at a maximal rate.

\bibliographystyle{apj} 
\bibliography{mwmdf}

\begin{thebibliography}{52}
\expandafter\ifx\csname natexlab\endcsname\relax\def\natexlab#1{#1}\fi

\bibitem[{{Abadi} {et~al.}(2003){Abadi}, {Navarro}, {Steinmetz}, \&
  {Eke}}]{abadi2003}
{Abadi}, M.~G., {Navarro}, J.~F., {Steinmetz}, M., \& {Eke}, V.~R. 2003, \apj,
  597, 21

\bibitem[{{Alam} {et~al.}(2015){Alam}, {Albareti}, {Allende Prieto}, {Anders},
  {Anderson}, {Anderton}, {Andrews}, {Armengaud}, {Aubourg}, {Bailey}, \&
  et~al.}]{alam2015}
{Alam}, S., {Albareti}, F.~D., {Allende Prieto}, C., {et~al.} 2015, \apjs, 219,
  12

\bibitem[{{Andrews} {et~al.}(2016){Andrews}, {Weinberg}, {Sch{\"o}nrich}, \&
  {Johnson}}]{andrews2016}
{Andrews}, B.~H., {Weinberg}, D.~H., {Sch{\"o}nrich}, R., \& {Johnson}, J.~A.
  2016, ArXiv e-prints

\bibitem[{{Bensby} {et~al.}(2003){Bensby}, {Feltzing}, \&
  {Lundstr{\"o}m}}]{bensby2003}
{Bensby}, T., {Feltzing}, S., \& {Lundstr{\"o}m}, I. 2003, \aap, 410, 527

\bibitem[{{Bird} {et~al.}(2013){Bird}, {Kazantzidis}, {Weinberg}, {Guedes},
  {Callegari}, {Mayer}, \& {Madau}}]{bird2013}
{Bird}, J.~C., {Kazantzidis}, S., {Weinberg}, D.~H., {et~al.} 2013, \apj, 773,
  43

\bibitem[{{Bournaud} {et~al.}(2009){Bournaud}, {Elmegreen}, \&
  {Martig}}]{bournaud2009}
{Bournaud}, F., {Elmegreen}, B.~G., \& {Martig}, M. 2009, \apjl, 707, L1

\bibitem[{{Bovy} {et~al.}(2012){Bovy}, {Rix}, \& {Hogg}}]{bovy2012a}
{Bovy}, J., {Rix}, H.-W., \& {Hogg}, D.~W. 2012, \apj, 751, 131

\bibitem[{{Bressan} {et~al.}(2012){Bressan}, {Marigo}, {Girardi}, {Salasnich},
  {Dal Cero}, {Rubele}, \& {Nanni}}]{bressan2012}
{Bressan}, A., {Marigo}, P., {Girardi}, L., {et~al.} 2012, \mnras, 427, 127

\bibitem[{{Brook} {et~al.}(2004){Brook}, {Kawata}, {Gibson}, \&
  {Freeman}}]{brook2004}
{Brook}, C.~B., {Kawata}, D., {Gibson}, B.~K., \& {Freeman}, K.~C. 2004, \apj,
  612, 894

\bibitem[{{Chiappini} {et~al.}(1997){Chiappini}, {Matteucci}, \&
  {Gratton}}]{chiappini1997}
{Chiappini}, C., {Matteucci}, F., \& {Gratton}, R. 1997, \apj, 477, 765

\bibitem[{{Chieffi} \& {Limongi}(2004)}]{chieffi2004}
{Chieffi}, A., \& {Limongi}, M. 2004, \apj, 608, 405

\bibitem[{{Dav{\'e}} {et~al.}(2012){Dav{\'e}}, {Finlator}, \&
  {Oppenheimer}}]{dave2012}
{Dav{\'e}}, R., {Finlator}, K., \& {Oppenheimer}, B.~D. 2012, \mnras, 421, 98

\bibitem[{{De Silva} {et~al.}(2015){De Silva}, {Freeman}, {Bland-Hawthorn},
  {Martell}, {de Boer}, {Asplund}, {Keller}, {Sharma}, {Zucker}, {Zwitter},
  {Anguiano}, {Bacigalupo}, {Bayliss}, {Beavis}, {Bergemann}, {Campbell},
  {Cannon}, {Carollo}, {Casagrande}, {Casey}, {Da Costa}, {D'Orazi}, {Dotter},
  {Duong}, {Heger}, {Ireland}, {Kafle}, {Kos}, {Lattanzio}, {Lewis}, {Lin},
  {Lind}, {Munari}, {Nataf}, {O'Toole}, {Parker}, {Reid}, {Schlesinger},
  {Sheinis}, {Simpson}, {Stello}, {Ting}, {Traven}, {Watson}, {Wittenmyer},
  {Yong}, \& {{\v Z}erjal}}]{desilva2015}
{De Silva}, G.~M., {Freeman}, K.~C., {Bland-Hawthorn}, J., {et~al.} 2015,
  \mnras, 449, 2604

\bibitem[{{Edvardsson} {et~al.}(1993){Edvardsson}, {Andersen}, {Gustafsson},
  {Lambert}, {Nissen}, \& {Tomkin}}]{edvardsson1993}
{Edvardsson}, B., {Andersen}, J., {Gustafsson}, B., {et~al.} 1993, \aap, 275,
  101

\bibitem[{{Eisenstein} {et~al.}(2011){Eisenstein}, {Weinberg}, {Agol},
  {Aihara}, {Allende Prieto}, {Anderson}, {Arns}, {Aubourg}, {Bailey},
  {Balbinot}, \& et~al.}]{eisenstein2011}
{Eisenstein}, D.~J., {Weinberg}, D.~H., {Agol}, E., {et~al.} 2011, \aj, 142, 72

\bibitem[{{Finlator} \& {Dav{\'e}}(2008)}]{finlator2008}
{Finlator}, K., \& {Dav{\'e}}, R. 2008, \mnras, 385, 2181

\bibitem[{{Forbes} {et~al.}(2012){Forbes}, {Krumholz}, \&
  {Burkert}}]{forbes2012}
{Forbes}, J., {Krumholz}, M., \& {Burkert}, A. 2012, \apj, 754, 48

\bibitem[{{Forbes} {et~al.}(2014){Forbes}, {Krumholz}, {Burkert}, \&
  {Dekel}}]{forbes2014}
{Forbes}, J.~C., {Krumholz}, M.~R., {Burkert}, A., \& {Dekel}, A. 2014, \mnras,
  438, 1552

\bibitem[{{Garc{\'{\i}}a P{\'e}rez} {et~al.}(2016){Garc{\'{\i}}a P{\'e}rez},
  {Allende Prieto}, {Holtzman}, {Shetrone}, {M{\'e}sz{\'a}ros}, {Bizyaev},
  {Carrera}, {Cunha}, {Garc{\'{\i}}a-Hern{\'a}ndez}, {Johnson}, {Majewski},
  {Nidever}, {Schiavon}, {Shane}, {Smith}, {Sobeck}, {Troup}, {Zamora},
  {Weinberg}, {Bovy}, {Eisenstein}, {Feuillet}, {Frinchaboy}, {Hayden},
  {Hearty}, {Nguyen}, {O'Connell}, {Pinsonneault}, {Wilson}, \&
  {Zasowski}}]{garcia-perez2016}
{Garc{\'{\i}}a P{\'e}rez}, A.~E., {Allende Prieto}, C., {Holtzman}, J.~A.,
  {et~al.} 2016, \aj, 151, 144

\bibitem[{{Gilmore} \& {Reid}(1983)}]{gilmore1983}
{Gilmore}, G., \& {Reid}, N. 1983, \mnras, 202, 1025

\bibitem[{{Gilmore} {et~al.}(2012){Gilmore}, {Randich}, {Asplund}, {Binney},
  {Bonifacio}, {Drew}, {Feltzing}, {Ferguson}, {Jeffries}, {Micela},
  {Negueruela}, {Prusti}, {Rix}, {Vallenari}, {Alfaro}, {Allende-Prieto},
  {Babusiaux}, {Bensby}, {Blomme}, {Bragaglia}, {Flaccomio}, {Fran{\c c}ois},
  {Irwin}, {Koposov}, {Korn}, {Lanzafame}, {Pancino}, {Paunzen},
  {Recio-Blanco}, {Sacco}, {Smiljanic}, {Van Eck}, \& {Walton}}]{gilmore2012}
{Gilmore}, G., {Randich}, S., {Asplund}, M., {et~al.} 2012, The Messenger, 147,
  25

\bibitem[{{H{\"a}nninen} \& {Flynn}(2002)}]{hanninen2002}
{H{\"a}nninen}, J., \& {Flynn}, C. 2002, \mnras, 337, 731

\bibitem[{{Hayden} {et~al.}(2015){Hayden}, {Bovy}, {Holtzman}, {Nidever},
  {Bird}, {Weinberg}, {Andrews}, {Majewski}, {Allende Prieto}, {Anders},
  {Beers}, {Bizyaev}, {Chiappini}, {Cunha}, {Frinchaboy},
  {Garc{\'{\i}}a-Her{\'n}andez}, {Garc{\'{\i}}a P{\'e}rez}, {Girardi},
  {Harding}, {Hearty}, {Johnson}, {M{\'e}sz{\'a}ros}, {Minchev}, {O'Connell},
  {Pan}, {Robin}, {Schiavon}, {Schneider}, {Schultheis}, {Shetrone},
  {Skrutskie}, {Steinmetz}, {Smith}, {Wilson}, {Zamora}, \&
  {Zasowski}}]{hayden2015}
{Hayden}, M.~R., {Bovy}, J., {Holtzman}, J.~A., {et~al.} 2015, \apj, 808, 132

\bibitem[{{Holtzman} {et~al.}(2015){Holtzman}, {Shetrone}, {Johnson}, {Allende
  Prieto}, {Anders}, {Andrews}, {Beers}, {Bizyaev}, {Blanton}, {Bovy},
  {Carrera}, {Chojnowski}, {Cunha}, {Eisenstein}, {Feuillet}, {Frinchaboy},
  {Galbraith-Frew}, {Garc{\'{\i}}a P{\'e}rez}, {Garc{\'{\i}}a-Hern{\'a}ndez},
  {Hasselquist}, {Hayden}, {Hearty}, {Ivans}, {Majewski}, {Martell},
  {Meszaros}, {Muna}, {Nidever}, {Nguyen}, {O'Connell}, {Pan}, {Pinsonneault},
  {Robin}, {Schiavon}, {Shane}, {Sobeck}, {Smith}, {Troup}, {Weinberg},
  {Wilson}, {Wood-Vasey}, {Zamora}, \& {Zasowski}}]{holtzman2015}
{Holtzman}, J.~A., {Shetrone}, M., {Johnson}, J.~A., {et~al.} 2015, \aj, 150,
  148

\bibitem[{{Iwamoto} {et~al.}(1999){Iwamoto}, {Brachwitz}, {Nomoto},
  {Kishimoto}, {Umeda}, {Hix}, \& {Thielemann}}]{iwamoto1999}
{Iwamoto}, K., {Brachwitz}, F., {Nomoto}, K., {et~al.} 1999, \apjs, 125, 439

\bibitem[{{Juri{\'c}} {et~al.}(2008){Juri{\'c}}, {Ivezi{\'c}}, {Brooks},
  {Lupton}, {Schlegel}, {Finkbeiner}, {Padmanabhan}, {Bond}, {Sesar},
  {Rockosi}, {Knapp}, {Gunn}, {Sumi}, {Schneider}, {Barentine}, {Brewington},
  {Brinkmann}, {Fukugita}, {Harvanek}, {Kleinman}, {Krzesinski}, {Long},
  {Neilsen}, {Nitta}, {Snedden}, \& {York}}]{juric2008}
{Juri{\'c}}, M., {Ivezi{\'c}}, {\v Z}., {Brooks}, A., {et~al.} 2008, \apj, 673,
  864

\bibitem[{{Kalirai} {et~al.}(2008){Kalirai}, {Hansen}, {Kelson}, {Reitzel},
  {Rich}, \& {Richer}}]{kalirai2008}
{Kalirai}, J.~S., {Hansen}, B.~M.~S., {Kelson}, D.~D., {et~al.} 2008, \apj,
  676, 594

\bibitem[{{Kroupa}(2001)}]{kroupa2001}
{Kroupa}, P. 2001, \mnras, 322, 231

\bibitem[{{Lee} {et~al.}(2011){Lee}, {Beers}, {An}, {Ivezi{\'c}}, {Just},
  {Rockosi}, {Morrison}, {Johnson}, {Sch{\"o}nrich}, {Bird}, {Yanny},
  {Harding}, \& {Rocha-Pinto}}]{lee2011}
{Lee}, Y.~S., {Beers}, T.~C., {An}, D., {et~al.} 2011, \apj, 738, 187

\bibitem[{{Leroy} {et~al.}(2008){Leroy}, {Walter}, {Brinks}, {Bigiel}, {de
  Blok}, {Madore}, \& {Thornley}}]{leroy2008}
{Leroy}, A.~K., {Walter}, F., {Brinks}, E., {et~al.} 2008, \aj, 136, 2782

\bibitem[{{Limongi} \& {Chieffi}(2006)}]{limongi2006}
{Limongi}, M., \& {Chieffi}, A. 2006, \apj, 647, 483

\bibitem[{{Luo} {et~al.}(2015){Luo}, {Zhao}, {Zhao}, {Deng}, {Liu}, {Jing},
  {Wang}, {Zhang}, {Shi}, {Cui}, {Chu}, {Li}, {Bai}, {Wu}, {Cai}, {Cao}, {Cao},
  {Carlin}, {Chen}, {Chen}, {Chen}, {Chen}, {Chen}, {Chen}, {Chen},
  {Christlieb}, {Chu}, {Cui}, {Dong}, {Du}, {Fan}, {Feng}, {Fu}, {Gao}, {Gong},
  {Gu}, {Guo}, {Han}, {He}, {Hou}, {Hou}, {Hou}, {Hu}, {Hu}, {Hu}, {Huo},
  {Jia}, {Jiang}, {Jiang}, {Jiang}, {Jin}, {Kong}, {Kong}, {Lei}, {Li}, {Li},
  {Li}, {Li}, {Li}, {Li}, {Li}, {Li}, {Li}, {Li}, {Li}, {Li}, {Liang}, {Lin},
  {Liu}, {Liu}, {Liu}, {Liu}, {Lu}, {Luo}, {Mao}, {Newberg}, {Ni}, {Qi}, {Qi},
  {Shen}, {Shi}, {Song}, {Song}, {Su}, {Su}, {Tang}, {Tao}, {Tian}, {Wang},
  {Wang}, {Wang}, {Wang}, {Wang}, {Wang}, {Wang}, {Wang}, {Wang}, {Wang},
  {Wang}, {Wang}, {Wang}, {Wang}, {Wang}, {Wang}, {Wang}, {Wang}, {Wang},
  {Wang}, {Wei}, {Wei}, {Wu}, {Wu}, {Wu}, {Wu}, {Xing}, {Xu}, {Xu}, {Xu},
  {Yan}, {Yang}, {Yang}, {Yang}, {Yang}, {Yao}, {Yu}, {Yuan}, {Yuan}, {Yuan},
  {Yuan}, {Zhai}, {Zhang}, {Zhang}, {Zhang}, {Zhang}, {Zhang}, {Zhang},
  {Zhang}, {Zhang}, {Zhao}, {Zhou}, {Zhou}, {Zhu}, {Zhu}, {Zou}, \&
  {Zuo}}]{luo2015}
{Luo}, A.-L., {Zhao}, Y.-H., {Zhao}, G., {et~al.} 2015, Research in Astronomy
  and Astrophysics, 15, 1095

\bibitem[{{Majewski} {et~al.}(2016){Majewski}, {Schiavon}, {Frinchaboy},
  {Allende Prieto}, {Barkhouser}, {Bizyaev}, {Blank}, {Brunner}, {Burton},
  {Carrera}, {Chojnowski}, {Cunha}, {Epstein}, {Fitzgerald}, {Garcia Perez},
  {Hearty}, {Henderson}, {Holtzman}, {Johnson}, {Lam}, {Lawler}, {Maseman},
  {Meszaros}, {Nelson}, {Coung Nguyen}, {Nidever}, {Pinsonneault}, {Shetrone},
  {Smee}, {Smith}, {Stolberg}, {Skrutskie}, {Walker}, {Wilson}, {Zasowski},
  {Anders}, {Basu}, {Beland}, {Blanton}, {Bovy}, {Brownstein}, {Carlberg},
  {Chaplin}, {Chiappini}, {Eisenstein}, {Elsworth}, {Feuillet}, {Fleming},
  {Galbraith-Frew}, {Garcia}, {Anibal Garcia-Hernandez}, {Gillespie},
  {Girardi}, {Gunn}, {Hasselquist}, {Hayden}, {Hekker}, {Ivans}, {Kinemuchi},
  {Klaene}, {Mahadevan}, {Mathur}, {Mosser}, {Muna}, {Munn}, {Nichol},
  {O'Connell}, {Robin}, {Rocha-Pinto}, {Schultheis}, {Serenelli}, {Shane},
  {Silva Aguirre}, {Sobeck}, {Thompson}, {Troup}, {Weinberg}, \&
  {Zamora}}]{majewski2016}
{Majewski}, S.~R., {Schiavon}, R.~P., {Frinchaboy}, P.~M., {et~al.} 2016, ArXiv
  e-prints

\bibitem[{{Maoz} {et~al.}(2012){Maoz}, {Mannucci}, \& {Brandt}}]{maoz2012b}
{Maoz}, D., {Mannucci}, F., \& {Brandt}, T.~D. 2012, \mnras, 426, 3282

\bibitem[{{Martig} {et~al.}(2016){Martig}, {Fouesneau}, {Rix}, {Ness},
  {M{\'e}sz{\'a}ros}, {Garc{\'{\i}}a-Hern{\'a}ndez}, {Pinsonneault},
  {Serenelli}, {Silva Aguirre}, \& {Zamora}}]{martig2016}
{Martig}, M., {Fouesneau}, M., {Rix}, H.-W., {et~al.} 2016, \mnras, 456, 3655

\bibitem[{{Ness} {et~al.}(2016){Ness}, {Hogg}, {Rix}, {Martig}, {Pinsonneault},
  \& {Ho}}]{ness2016}
{Ness}, M., {Hogg}, D.~W., {Rix}, H.-W., {et~al.} 2016, \apj, 823, 114

\bibitem[{{Nidever} {et~al.}(2014){Nidever}, {Bovy}, {Bird}, {Andrews},
  {Hayden}, {Holtzman}, {Majewski}, {Smith}, {Robin}, {Garc{\'{\i}}a
  P{\'e}rez}, {Cunha}, {Allende Prieto}, {Zasowski}, {Schiavon}, {Johnson},
  {Weinberg}, {Feuillet}, {Schneider}, {Shetrone}, {Sobeck},
  {Garc{\'{\i}}a-Hern{\'a}ndez}, {Zamora}, {Rix}, {Beers}, {Wilson},
  {O'Connell}, {Minchev}, {Chiappini}, {Anders}, {Bizyaev}, {Brewington},
  {Ebelke}, {Frinchaboy}, {Ge}, {Kinemuchi}, {Malanushenko}, {Malanushenko},
  {Marchante}, {M{\'e}sz{\'a}ros}, {Oravetz}, {Pan}, {Simmons}, \&
  {Skrutskie}}]{nidever2014}
{Nidever}, D.~L., {Bovy}, J., {Bird}, J.~C., {et~al.} 2014, \apj, 796, 38

\bibitem[{{Nidever} {et~al.}(2015){Nidever}, {Holtzman}, {Allende Prieto},
  {Beland}, {Bender}, {Bizyaev}, {Burton}, {Desphande}, {Fleming},
  {Garc{\'{\i}}a P{\'e}rez}, {Hearty}, {Majewski}, {M{\'e}sz{\'a}ros}, {Muna},
  {Nguyen}, {Schiavon}, {Shetrone}, {Skrutskie}, {Sobeck}, \&
  {Wilson}}]{nidever2015}
{Nidever}, D.~L., {Holtzman}, J.~A., {Allende Prieto}, C., {et~al.} 2015, \aj,
  150, 173

\bibitem[{{Peeples} \& {Shankar}(2011)}]{peeples2011}
{Peeples}, M.~S., \& {Shankar}, F. 2011, \mnras, 417, 2962

\bibitem[{{Pinsonneault} {et~al.}(2014){Pinsonneault}, {Elsworth}, {Epstein},
  {Hekker}, {M{\'e}sz{\'a}ros}, {Chaplin}, {Johnson}, {Garc{\'{\i}}a},
  {Holtzman}, {Mathur}, {Garc{\'{\i}}a P{\'e}rez}, {Silva Aguirre}, {Girardi},
  {Basu}, {Shetrone}, {Stello}, {Allende Prieto}, {An}, {Beck}, {Beers},
  {Bizyaev}, {Bloemen}, {Bovy}, {Cunha}, {De Ridder}, {Frinchaboy},
  {Garc{\'{\i}}a-Hern{\'a}ndez}, {Gilliland}, {Harding}, {Hearty}, {Huber},
  {Ivans}, {Kallinger}, {Majewski}, {Metcalfe}, {Miglio}, {Mosser}, {Muna},
  {Nidever}, {Schneider}, {Serenelli}, {Smith}, {Tayar}, {Zamora}, \&
  {Zasowski}}]{pinsonneault2014}
{Pinsonneault}, M.~H., {Elsworth}, Y., {Epstein}, C., {et~al.} 2014, \apjs,
  215, 19

\bibitem[{{Sch{\"o}nrich} \& {Binney}(2009{\natexlab{a}})}]{schoenrich2009a}
{Sch{\"o}nrich}, R., \& {Binney}, J. 2009{\natexlab{a}}, \mnras, 396, 203

\bibitem[{{Sch{\"o}nrich} \& {Binney}(2009{\natexlab{b}})}]{schoenrich2009b}
---. 2009{\natexlab{b}}, \mnras, 399, 1145

\bibitem[{{Sellwood} \& {Binney}(2002)}]{sellwood2002}
{Sellwood}, J.~A., \& {Binney}, J.~J. 2002, \mnras, 336, 785

\bibitem[{{Simha} {et~al.}(2014){Simha}, {Weinberg}, {Conroy}, {Dave},
  {Fardal}, {Katz}, \& {Oppenheimer}}]{simha2014}
{Simha}, V., {Weinberg}, D.~H., {Conroy}, C., {et~al.} 2014, ArXiv e-prints

\bibitem[{{Stello} {et~al.}(2015){Stello}, {Huber}, {Sharma}, {Johnson},
  {Lund}, {Handberg}, {Buzasi}, {Silva Aguirre}, {Chaplin}, {Miglio},
  {Pinsonneault}, {Basu}, {Bedding}, {Bland-Hawthorn}, {Casagrande}, {Davies},
  {Elsworth}, {Garcia}, {Mathur}, {Di Mauro}, {Mosser}, {Schneider},
  {Serenelli}, \& {Valentini}}]{stello2015}
{Stello}, D., {Huber}, D., {Sharma}, S., {et~al.} 2015, \apjl, 809, L3

\bibitem[{{Talbot} \& {Arnett}(1971)}]{talbot1971}
{Talbot}, Jr., R.~J., \& {Arnett}, W.~D. 1971, \apj, 170, 409

\bibitem[{{Tinsley}(1980)}]{tinsley1980}
{Tinsley}, B.~M. 1980, \fcp, 5, 287

\bibitem[{{Villalobos} \& {Helmi}(2008)}]{villalobos2008}
{Villalobos}, {\'A}., \& {Helmi}, A. 2008, \mnras, 391, 1806

\bibitem[{{Vincenzo} {et~al.}(2016){Vincenzo}, {Matteucci}, \&
  {Spitoni}}]{vincenzo2016}
{Vincenzo}, F., {Matteucci}, F., \& {Spitoni}, E. 2016, ArXiv e-prints

\bibitem[{{Weinberg} {et~al.}(2016){Weinberg}, {Andrews}, \&
  {Freudenburg}}]{weinberg2016}
{Weinberg}, D.~H., {Andrews}, B.~H., \& {Freudenburg}, J. 2016, ArXiv e-prints

\bibitem[{{Wielen} {et~al.}(1996){Wielen}, {Fuchs}, \& {Dettbarn}}]{wielen1996}
{Wielen}, R., {Fuchs}, B., \& {Dettbarn}, C. 1996, \aap, 314

\bibitem[{{Zasowski} {et~al.}(2013){Zasowski}, {Johnson}, {Frinchaboy},
  {Majewski}, {Nidever}, {Rocha Pinto}, {Girardi}, {Andrews}, {Chojnowski},
  {Cudworth}, {Jackson}, {Munn}, {Skrutskie}, {Beaton}, {Blake}, {Covey},
  {Deshpande}, {Epstein}, {Fabbian}, {Fleming}, {Garcia Hernandez}, {Herrero},
  {Mahadevan}, {M{\'e}sz{\'a}ros}, {Schultheis}, {Sellgren}, {Terrien}, {van
  Saders}, {Allende Prieto}, {Bizyaev}, {Burton}, {Cunha}, {da Costa},
  {Hasselquist}, {Hearty}, {Holtzman}, {Garc{\'{\i}}a P{\'e}rez}, {Maia},
  {O'Connell}, {O'Donnell}, {Pinsonneault}, {Santiago}, {Schiavon}, {Shetrone},
  {Smith}, \& {Wilson}}]{zasowski2013}
{Zasowski}, G., {Johnson}, J.~A., {Frinchaboy}, P.~M., {et~al.} 2013, \aj, 146,
  81

\end{thebibliography}

\newpage

\end{document}